# Systematic study of $\alpha$-decay half-lives of superheavy nuclei based on Coulomb and proximity potential models with temperature effects

Panpan Qi [1], Xuanpeng Xiao [1], Gongming Yu [1], Haitao Yang [2], and Qiang Hu [3]

[1] College of Physics and Technology, Kunming University, Kunming 650214, China
[2] College of Science, Zhaotong University, Zhaotong 657000, China
[3] Institute of Modern Physics, Chinese Academy of Sciences, Lanzhou 730000, China

By employing the Coulomb proximity potential model (CPPM) in conjunction with 22 distinct proximity potential models, we investigated the temperature dependence and the effects of proton number and neutron number on the diffusion parameters that determine the $\alpha$-decay half-lives of superheavy nuclei. The results indicate that the Prox.77-3 T-DEP proximity potential model yields the best performance, with the lowest root mean square deviation ($\sigma$ = 0.515), reflecting a high consistency with experimental data. In contrast, Bass77, AW95, Ngô80, and Guo2013 display larger deviations. The inclusion of temperature dependence significantly improves the accuracy of models such as Prox.77-3, Prox.77-6, and Prox.77-7. The $\alpha$-decay half-lives of 36 potential superheavy nuclei were further predicted using the five most accurate proximity potential models and Ni's empirical formula, with the results aligning well with experimental data. These predictions underscore the high reliability of the CPPM combined with proximity potential models in the theoretical calculation of $\alpha$-decay half-lives of superheavy nuclei, offering valuable theoretical insights for future experimental investigations of superheavy nuclei.

## I. INTRODUCTION

Shortly after Rutherford and Geiger discovered $\alpha$-decay [1], Gamow explained it in 1928 as a quantum tunneling process [2]. Since then, $\alpha$-decay has remained a central topic in nuclear physics, being the primary decay mode for superheavy nuclei (SHN). In comparison to other decay modes, such as cluster decay and fission, it is considered a relatively straightforward process [3,4]. $\alpha$-decay is crucial for understanding the nuclear structure and stability of heavy and superheavy nuclei [5]. Moreover, it plays a significant role in the identification of new SHNs [6] and in the study of nuclear forces [7].

Numerous theoretical and experimental studies have been conducted on $\alpha$-decay. From a theoretical perspective, research efforts focus on developing suitable formalisms for calculating the $\alpha$-decay half-life, such as the unified Fission Model (UFM) [8], Relativistic Mean Field (RMF) Theory [9], Generalized Liquid Drop Model

---

Email: ygmanan@kmu.edu.cn, yanghaitao205@163.com, qianghu@impcas.ac.cn

(GLDM) [10], modified Generalized Liquid Drop Model(MGLDM) [11], Analytical Super Asymmetric Fission (ASAF) Model [12], Coulomb and Proximity Potential Model (CPPM) [13], and Density-Dependent Cluster Model (DDCM) [14]. The experimental identification of new elements through $\alpha$-decay is commonly employed, as $\alpha$-decay is the predominant decay mode in superheavy nuclei (SHN), which are synthesized via hot, warm, and cold fusion reactions [15,16]. The study of these nuclei is a key area of interest in nuclear physics, contributing to the understanding of concepts such as the island of stability, magic numbers, spin-parity, and deformed nuclei.

In the 1970s, the concept of proximity potential to handle heavy-ion reactions was first introduced by Blocki [17]. The nuclear proximity potential primarily consists of two components: one is a geometric function that depends on the shape and configuration of the interacting nuclei, while the other is a universal function $\Phi(s)$, which describes the short-range interaction as a function of the separation distance 's' between the two nuclei involved in the reaction [18]. The proximity potential exists in several versions, each with distinct characteristics compared to the original form (Prox.1977) [17]. These versions improve upon aspects such as the surface energy coefficients [19-21], the universal function and the nuclear radius parameterization [18,22,23]. These modifications have been utilized in various fields for comparative studies conducted by nuclear physicists [24-26]. It has also been extensively researched in the field of $\alpha$-decay half-lives of superheavy nuclei[27,28]. In 2019, the half lives of $\alpha$-decay for 70 superheavy nuclei (SHN) in the atomic number range $106 \leq Z \leq 118$ was estimated by Ghodsi [27] in 28 different versions of proximity potentials as well as a deformed method for the Coulomb potential. It was found that the values computed by Ngô (1980) are in good agreement with experimental half-lives, in comparison with other versions [27]. The detailed study of the $\alpha$-decay properties of isotopes of superheavy nuclei with $Z = 118$, covering the mass range $271 \leq A \leq 310$, was conducted by K. P. Santhosh [28]. in the Coulomb and proximity potential model for deformed nuclei. The $\alpha$-decay half-lives of $^{294}$118 and its decay products, calculated using their formalism, show a good agreement with experimental data [28].

The diffuseness parameter is crucial in the proximity potential, as it significantly influences decay characteristics. Various factors, including assault frequency, turning points, potential barriers, tunneling probability, and half-life, explicitly depend on the surface diffuseness parameter. Even a small variation in this parameter can result in substantial changes in these values [29]. In 2019, the diffuseness parameter of the universal proximity potential was examined by Aladdin Abdul-Latif and Omar Nagib [30]. They subsequently proposed an empirical method for calculating this parameter, which depends on the charge and neutron numbers. This study aims to analyze the diffusion parameters proposed by Aladdin Abdul-Latif and Omar Nagib and apply them to different proximity potentials, considering both cases where the proximity potential depends on temperature and where it does not. The CPPM model is employed to calculate the half-lives of superheavy nuclei. The results were compared with experimental values and those calculated using Ni's empirical formula.

The remainder of this article is organized as follows. Section II provides a detailed exposition of the theoretical framework of the CPPM, encompassing 22 different proximity potential formalisms (including both temperature-dependent and temperature-independent cases), along with a comparative analysis of Ni's empirical formula. Additionally, the diffusion parameters proposed by Aladdin Abdul-Latif and Omar Nagib are introduced. Section III presents a systematic analysis and discussion of the results. Finally, Section IV offers a concise summary of the entire work.

## II. GENERAL FORMALISM
### A. Half-lives of $\alpha$-decay

The half-life of superheavy nuclei can be determined as follows [30]:

$$T_{1/2} = \frac{\pi \hbar \ln 2}{P_0 E_v}[1 + \exp(B)], \#(1)$$

where $P_0$, $E_v$, and $B$ represent the preformation factor, the zero-point vibration energy, and the action integral, respectively. The action integral is given by

$$B = \frac{2}{\hbar}\int_{R_{\text{in}}}^{R_{\text{out}}} \sqrt{2\mu(V(r) - Q_\alpha)}\, dr, \#(2)$$

where $Q_\alpha$, $\mu$, $R_{\text{in}}$ and $R_{\text{out}}$ are the decay energy, reduced mass, and first and second turning points, respectively. The zero-point vibration energy is given by

$$E_v = \frac{\hbar\omega}{2} = \frac{\hbar\pi}{2R_p}\sqrt{\frac{2E_\alpha}{m_\alpha}}\#(3)$$

In this formula, $R_p$, $E_\alpha$, $m_\alpha$ represent the parent radius, the kinetic energy of the $\alpha$-nuclei, and its mass, respectively. The system's decay energy $Q_\alpha$ and the kinetic energy of $\alpha$-nuclei can be related by the following equation [31]:

$$Q_\alpha = \frac{A_d}{A_p}E_\alpha + \left[6.53Z_d^{7/5} - 8.0Z_d^{2/5}\right] \times 10^{-5}\text{ MeV}.\#(4)$$

Here, $A_d$ and $A_p$ are the mass numbers of the daughter and parent nuclei, respectively, and $Z_d$ is the proton number of the daughter nucleus. In this study, the preformation factor is characterized by [32]

$$\log_{10} P_o = a + b(Z - Z_1)(Z_2 - Z_1) + c(N - N_1)(N_2 - N)$$
$$+ dA + e(Z - Z_1)(N - N_1).\#(5)$$

The values of $a$, $b$, $c$, $d$, and $e$ are 34.90593, 0.003011, 0.003717, 0.151216, and 0.006681, respectively. The magic numbers $Z_1$, $Z_2$, $N_1$, and $N_2$ are 82, 126, 152, and 184, respectively. The total interaction potential between the emitted $\alpha$ particle and the daughter nucleus includes the nuclear potential, Coulomb potential, and centrifugal potential. It can be expressed as follows:

$$V(r) = V_N(r) + V_C(r) + V_l(r). \#(6)$$

Where $r$ is the separation distance between the center of the daughter nucleus and the $\alpha$ particle. In this study, we adopt the proximity potential form to replace the nuclear potential. Additionally, we consider the case where the total angular momentum is not carried in the $\alpha$-daughter system, and therefore $V_l(r)$ is neglected. The Coulomb potential $V_c(r)$ is taken as the potential of a uniformly charged sphere with radius $R$, and it can be expressed as follows:

$$V_c(r) = \begin{cases} \dfrac{Z_\alpha Z_d e^2}{2R}\left[3 - \left(\dfrac{r}{R}\right)^2\right], & r \leq R, \\ \dfrac{Z_\alpha Z_d e^2}{r}, & r > R. \end{cases} \quad \#(7)$$

The radius $R = R_1 + R_2$ represent the combined radii of the daughter nucleus and the emitted α particle, respectively. different expressions for $R_i$ (where $i = 1,2$) are given under different forms of the potential. $Z_1$ and $Z_2$ denote the proton numbers of the daughter nucleus and the α particle, respectively.

B. Proximity potential formalism

We extend the CPPM to investigate the half-lives of α-decay using 22 distinct versions of proximity potential formalisms. These include: (i) Prox.77 [17] and its 12 modified variants, which are based on adjustments to the surface energy coefficient $\gamma_0$ and $k_s$ [19-21,33-38]; (ii) Prox.81 [22]; (iii) Bass73 [23,39] and its revised versions, Bass77 [40] and Bass80 [41]; (iv) CW76 [42] and its revised version BW91 [41] and AW95 [43]; (v) Ngô80 [44]; and (vi) Guo2013 [45]. The detailed expressions for these proximity potential formalisms are presented below.

1. Proximity potential 77 family

The original form of the proximity potential for two spherical interacting nuclei was introduced by Blocki [17] and can be written as follows:
$$V_N(r) = 4\pi\gamma b \overline{R} \Phi(\xi). \#(8)$$
Here, γ represents the surface energy coefficient, which is derived from the formula by Myers and swiatecki [33],
$$\gamma = \gamma_0(1 - k_s I^2), \#(9)$$
where $I = \dfrac{N_p - Z_p}{N_p + Z_p}$ represents the asymmetry parameter, which indicates the neutron-proton imbalance of the parent nucleus, with $N_p$ and $Z_p$ being the neutron and proton numbers of the parent nucleus, respectively. The surface energy constant $\gamma_0$ and the surface asymmetry constant $k_s$ for Prox.77, along with its modifications, are provided in Table 1. The mean curvature radius, denoted as $\overline{R}$, can be calculated by
$$\overline{R} = \frac{C_\alpha C_d}{C_\alpha + C_d}. \#(10)$$
The quantity $C_i$ is given by the expression $C_i = R_i\left[1 - \left(\dfrac{b}{R_i}\right)^2\right]$ $(i = \alpha, d)$, with $C_\alpha$ and $C_d$ representing the matter radii of daughter nucleus and the α particle, respectively. The effective sharp radius $R_i$ is expressed by the equation $R_i = 1.28 A_i^{1/3} - 0.76 + 0.8 A_i^{-1/3}$ $(i = \alpha, d)$, where $A_i$ denotes the mass number of the emitted α particle $(i = \alpha)$ and daughter nucleus $(i = d)$, respectively. The nuclear surface diffusion parameters used in this work were proposed by Aladdin Abdul-latif and Omar Nagib in 2019 [30], and they are expressed as follows:
$$b = \frac{\pi}{\sqrt{3}}(-1.09535 + 0.012063 Z_p + 0.0019759 N_p). \#(11)$$

The universal function $\Phi(\xi)$ takes the following form

$$\phi(\xi) = \begin{cases} \frac{1}{2}(\xi - 2.54)^2 - 0.0852(\xi - 2.54)^3, & \xi < 1.2511, \\ -3.437 exp\left(-\frac{\xi}{0.75}\right), & \xi \geq 1.2511. \end{cases} \quad \#(12)$$

Here, $\xi = \frac{r - C_\alpha - C_d}{b}$ represents the distance between the near surface of the daughter nucleus and the α particle.

Table1. The different sets of surface energy coefficients for Prox.77 and its modifications are presented. $\gamma_0$ and $k_s$ represent the surface energy constant and the surface asymmetry constant, respectively.

| γ set | $\gamma_0$(MeV/fm²) | $k_s$ | References |
|---|---|---|---|
| Set1(γ-MS 1967) | 0.9517 | 1.7826 | [33] |
| Set2(γ-MS 1966) | 1.01734 | 1.79 | [19] |
| Set3(γ-MS 1976) | 1.460734 | 4.0 | [20] |
| Set4(γ-KNS 1979) | 1.2402 | 3.0 | [21] |
| Set5(γ-MN-I 1981) | 1.1756 | 2.2 | [34] |
| Set6(γ-MN-II 1981) | 1.27326 | 2.5 | [34] |
| Set7(γ-MN-III 1981) | 1.2502 | 2.4 | [34] |
| Set8(γ-RR 1984) | 0.9517 | 2.6 | [35] |
| Set9(γ-MN 1988) | 1.2496 | 2.3 | [36] |
| Set10(γ-MN 1995) | 1.25284 | 2.345 | [37] |
| Set11(γ-PD-LDM 2003) | 1.08948 | 1.9830 | [38] |
| Set12(γ-PD-NLD 2003) | 0.9180 | 0.7546 | [38] |
| Set13(γ-PD-LSD 2003) | 0.911445 | 2.2938 | [38] |

2. Proximity potential Prox.81

In 1981, Blocki and Swiatecki introduced a new proximity potential formalism based on the proximity force theorem, referred to as Prox.81 [22]. This formalism differs from Prox.77 [17] in both the surface energy coefficient and the universal function, while the other aspects remain unchanged. The surface energy coefficient is given by $\gamma = 0.9517\left[1 - 1.7826\left(\frac{Np-Zp}{Np+Zp}\right)^2\right]$, and the universal function can be expressed as:

$$\Phi(\xi) = \begin{cases} -1.7817 + 0.9270\xi + 0.143\xi^2 - 0.09\xi^3, & \xi < 0, \\ -1.7817 + 0.9270\xi + 0.01696\xi^2 - 0.05148\xi^3, & 0 \leq \xi \leq 1.9475, \\ -4.41 exp\left(-\frac{\xi}{0.7176}\right), & \xi > 1.9475. \end{cases} \quad \#(13)$$

3. Proximity potential Bass73

In Ref [23,39], Bass derived an expression for the nuclear potential based on the liquid drop model. This expression is formulated as the difference between finite and infinite separation, denoted by $\xi$. The nuclear potential is given by

$$V_N(r) = -4\pi\gamma \frac{dR_\alpha R_d}{R}\exp\left(-\frac{\xi}{d}\right) = -\frac{da_s A_\alpha^{1/3} A_d^{1/3}}{R}\exp\left(-\frac{r-R}{d}\right). \#(14)$$

The variate $\gamma$ denotes the specific surface energy in the liquid drop model, while $d$ represents the range parameter, with a value of 1.35 fm. $a_s = 17.0$ MeV are the surface term in the liquid drop model mass formula. The radius $R$ is the sum of the half-maximum density radii, expressed as $R = R_\alpha + R_d = r_0(A_\alpha^{1/3} + A_d^{1/3})$, where $r_0 = 1.07$, $R_d$ and $R_\alpha$ are the radii of the daughter nucleus and the emitted $\alpha$ particle, respectively, and $A_d$ and $A_\alpha$ are their corresponding mass numbers.

4. Proximity potential Bass77

In 1977, R. Bass derived a universal nucleus-nucleus potential through a classical analysis of experimental fusion cross sections [40], which is expressed as

$$V_N(r) = -\frac{R_\alpha R_d}{R_\alpha + R_d}\Phi(\xi). \#(15)$$

Here, the variate $R_i = 1.16A_i^{1/3} - 1.39A_i^{1/3}(i = \alpha, d)$ refers to the half-density radius, where $A_i$ represents the mass number of the emitted $\alpha$ particle ($i = \alpha$) and daughter nucleus ($i = d$). The universal function $\Phi(\xi = r - R_\alpha - R_d)$ is given by

$$\Phi(s) = \left[0.03\exp\left(\frac{\xi}{3.3}\right) + 0.0061\exp\left(\frac{\xi}{0.65}\right)\right]^{-1}. \#(16)$$

5. Proximity potential Bass80

For the Bass80, the nuclear potential is given by [41]

$$V_N(r) = -\frac{R_\alpha R_d}{R_\alpha + R_d}\phi(r - R_\alpha - R_d). \#(17)$$

Here, $R_i = R_{si}\left(1 - \frac{0.98}{R_{si}^2}\right)$ with $R_{si} = 1.28A_i^{1/3} - 0.76 + 0.8A_i^{-1/3}$ ($i = \alpha, d$). The universal function $\phi(\xi = r - R_\alpha - R_d)$ has the form as

$$\Phi(s) = \left[0.033\exp\left(\frac{\xi}{3.5}\right) + 0.007\exp\left(\frac{\xi}{0.65}\right)\right]^{-1}. \#(18)$$

6. Proximity potential CW76

In Ref [42], Christensen and Winther analyzed the heavy-ion elastic scattering data and derived an empirical nuclear potential:

$$V_N(r) = -50\frac{R_\alpha R_d}{R_\alpha + R_d}\phi(r - R_\alpha - R_d), \#(19)$$

where $R_\alpha$ and $R_d$ are defined as follows:

$$R_i = 1.233A_i^{\frac{1}{3}} - 0.978A_i^{-\frac{1}{3}}, (i = \alpha, d). \#(20)$$

7. Proximity potential BW91

In 1991, Broglia and Winther introduced a more refined nuclear potential by adopting the Woods-Saxon parametrization of the proximity potential, originally formulated by CW76 [41]. This potential, referred to as BW91, is expressed:

$$V_N(r) = -\frac{V_0}{1+exp\left(\frac{r-R}{0.63}\right)} = -\frac{16\pi\gamma a \frac{R_\alpha R_d}{R_\alpha+R_d}}{1+exp\left(\frac{r-R}{0.63}\right)}. \#(21)$$

In the above formula, $\gamma$ represents the surface energy coefficient, which can be given by the following expression:

$$\gamma = 0.95\left[1-1.8\left(\frac{N_d-Z_d}{N_d+Z_d}\right)\left(\frac{N_\alpha-Z_\alpha}{N_\alpha+Z_\alpha}\right)\right]. \#(22)$$

Here, $N_d$ and $Z_d$ represent the numbers of neutrons and protons in the daughter nucleus, whereas $N_\alpha$ and $Z_\alpha$ are the numbers of neutrons and protons in the alpha particle. In this case, $a$ is assigned a value of 0.63 fm and $R = R_\alpha + R_d + 0.29$, where $R_i = 1.233 A_i^{1/3} - 0.98 A_i^{-1/3}$ $(i = \alpha, d)$.

8. Proximity potential AW95

The proximity potential expressions and parameters of AW95 and BW91 are essentially identical [43], although the values of the three parameters $a$, $R$, and $R_i (i = \alpha, d)$ differ. For AW95, the parameter $a$ is given by $\frac{1}{1.17\left(1+0.53\left(A_c^{-1/3}+A_d^{-1/3}\right)\right)}$, and $R$ is expressed as $R = R_\alpha + R_d$, where $R_i = 1.2 A_i^{1/3} - 0.09$ $(i = \alpha, d)$.

9. Proximity potential Ngô80

In 1980, H. Ngô and Ch. Ngô proposed a nuclear potential by calculating the real part of the interaction potential between two heavy ions within the framework of the sudden approximation [44]. This nuclear potential is referred to as Ngô80 and is mathematically represented as

$$V_N(r) = \frac{C_\alpha C_d}{C_\alpha+C_d}\phi(\xi). \#(23)$$

The matter radius of the alpha particle $C_\alpha$ and the daughter nucleus $C_d$, as well as the nuclear surface diffusion parameter $b$, are defined in the same form as in Prox77 [17]. $R_i$ denotes the sharp radii

$$R_i = \frac{N_i R_{ni} + Z_i R_{pi}}{A_i}, (i = c, d). \#(24)$$

where $R_{ji} = r_{0ji} A_i^{1/3}$ $(j = p, n, i = \alpha, d)$, with $r_{0pi} = 1.128$ fm and $r_{0ni} = 1.11375 + 1.875 \times 10^{-4} A_i$ fm. The universal function $\phi(\xi = r - C_\alpha - C_d)$ is expressed as

$$\Phi(\xi) = \begin{cases} -33 + 5.4(\xi+1.6)^2, & \xi < -1.6 \\ -33 exp\left(-\frac{1}{5}(\xi+1.6)^2\right), & \xi \geq -1.6. \end{cases} \#(25)$$

10. Proximity potential Guo2013

In Ref [45], Guo introduced a new universal function for the proximity potential model by using the double folding model with a density-dependent nucleon-nucleon interaction and fitting universal functions across various reaction systems. This proximity potential is referred to as Guo2013

$$V_N(r) = 4\pi\gamma b \frac{R_\alpha R_d}{R_\alpha + R_d} \phi(\xi). \#(26)$$

The nuclear surface diffusion parameter $b$, along with the effective sharp radius $R_i$ (where $i = \alpha, d$), is defined in the same form as in Prox77. The surface coefficient $\gamma$ can be obtained by

$$\gamma = 0.9517\left[1 - 1.7826\left(\frac{N_p - Z_p}{N_p + Z_p}\right)^2\right]. \#(27)$$

The universal function $\Phi\left(\xi = \frac{r - R_\alpha - R_d}{b}\right)$ is expressed as

$$\Phi(\xi) = \frac{p_1}{1 + exp\left(\frac{\xi + p_2}{p_3}\right)}, \#(28)$$

where $p_1 = -17.72$, $p_2 = 1.3$, $p_3 = 0.854$ are adjustable parameters.

### C. Temperature dependent proximity potential

In the previous sections, we studied the temperature-independent form of the proximity potential. In this section, we investigate thermal effects by using the temperature-dependent forms of the parameters $R$, $\gamma$, and $b$. These are given by [46-48]

$$R_i(T) = R_i(T = 0)[1 + 0.0005T^2] fm, (i = 1,2), \#(29)$$

$$\gamma(T) = \gamma(T = 0)\left[1 - \frac{T - T_b}{T_b}\right]^{\frac{2}{3}}, \#(30)$$

$$b(T) = b(T = 0)[1 + 0.009T^2]. \#(31)$$

where $T_b$ represents the temperature corresponding to energies near the Coulomb barrier. In this study, $b(T = 0)$ and $R_i(T = 0)$ represent the nuclear surface diffusion parameters and the normal radius formula, respectively, each associated with the proximity potential we have used. Additionally, an alternative form of the temperature-dependent surface energy coefficient is given by $\gamma(T) = \gamma(0)(1 - 0.07T)^2$ [49]. The temperature $T$ (in MeV) can be written as [50,51]:

$$E_p^* = E_{kin} + Q_{in} = \frac{1}{9}A_p T^2 - T. \#(32)$$

Where $E_p^*$ denotes the excitation energy of the parent nucleus, and $A_p$ represents its mass number. $Q_{in}$ is the entrance channel Q-value of the system. To calculate $E_{kin}$, the kinetic energy of the emitted particle, we use the following equation:

$$E_{kin} = \left(\frac{A_d}{A_p}\right)Q. \#(33)$$

### D. Ni's empirical formula

In 2008, Ni proposed a universal formula applicable to both α-decay and cluster radioactivity, establishing a relationship between the half-life and decay energy [52]. This formula was derived directly from an approximation of the WKB barrier penetration probability and can be expressed as [52]:

$$log_{10} T_{1/2} = a \sqrt{\mu}(Z_c Z_d Q_c^{-1/2}) + b \sqrt{\mu}(Z_c Z_d)^{1/2} + c. \#(34)$$

Where $Q_c$ represents the energy released during α-decay and cluster radioactivity processes, and μ is the reduced mass. In α-decay, the adjustable parameters are defined as ( $a$ = 0.39961 ), ( $b$ = 1.31008 ), and ( $c$ ), which take different values depending on the nuclear type: ($c$ = -17.00698) for even-even nuclei, ( $c$ = -16.26029) for even-odd nuclei, and ( $c$ = -16.40484) for odd-even nuclei [52].

### III. RESULTS DISCUSSION

The primary objective of this study is to incorporate a diffusion parameter related to the proton number (Z) and the neutron number (N) into 22 different proximity potentials, while also considering the temperature dependence of the proximity potential. The half-lives of superheavy nuclei were calculated using the Coulomb and Proximity Potential Model (CPPM). To determine the most suitable proximity potential form for describing α-decay in superheavy nuclei, we systematically computed the α-decay half-lives of 63 superheavy nuclei by employing 22 different versions of proximity potentials within the CPPM framework. Additionally, the Ni's empirical formula was utilized. The results are presented in detail in Table 2. In this table, the first through third columns represent the proton number, mass number, and neutron number of the parent nucleus, respectively. The fourth and fifth columns denote the energy released during the α-decay of the superheavy nucleus and the temperature of the parent nucleus, respectively. The last 11 columns present the experimental data for the α-decay half-lives of the superheavy nuclei, along with the calculated results obtained using the CPPM model with 22 different versions of proximity potentials (including temperature-independent proximity potentials, T-IND, and temperature-dependent proximity potentials, T-DEP), as well as Ni's empirical formula. From this table, it can be observed that the results by using Prox.77 and its 12 modified forms, as well as Prox.81, Bass73, Bass80, CW76, and BW91 (both T-IND and T-DEP), agree with the experimental data within one order of magnitude for most cases, except for a few superheavy nuclei decays. This indicates an excellent reproduction of the experimental data. In contrast, for Bass77 and AW95 (both T-IND and T-DEP), the results generally deviate from the experimental data by 1 to 2 orders of magnitude, while for Ngô80 and Guo2013 (both T-IND and T-DEP), the deviations range from 1 to 3 orders of magnitude. These findings highlight the varying accuracy of different proximity potentials in describing the α-decay half-lives of superheavy nuclei.

Table2. The half-lives of alpha decay for isotopes with Z = 84 − 117 were calculated using different formalisms. The $Q_\alpha$ values from experiments are taken from references [29,53].

| Parent Nuclei | | | $Q_\alpha(Mev)$ | Temp (Mev) | $log_{10}T_{1/2}/s$ | | | | | | | | | |
|---|---|---|---|---|---|---|---|---|---|---|---|---|---|---|
| | | | | | Expt | Prox.77-1 | | Prox.77-2 | | Prox.77-3 | | Prox.77-4 | | Prox.77-5 | |
| Z | A | N | | | | T-IND | T-DEP | T-IND | T-DEP | T-IND | T-DEP | T-IND | T-DEP | T-IND | T-DEP |
| 104 | 259 | 155 | 9.13 | 0.811 | 0.415 | 1.354 | 1.601 | 1.226 | 1.449 | 0.799 | 0.964 | 0.977 | 1.163 | 1.008 | 1.197 |
| 104 | 263 | 159 | 8.25 | 0.766 | 3.301 | 3.880 | 4.137 | 3.742 | 3.972 | 3.303 | 3.473 | 3.485 | 3.676 | 3.509 | 3.704 |
| 105 | 256 | 151 | 9.34 | 0.825 | 0.453 | 1.597 | 1.850 | 1.469 | 1.697 | 1.013 | 1.180 | 1.203 | 1.392 | 1.244 | 1.438 |
| 105 | 258 | 153 | 9.5 | 0.829 | 0.747 | 0.847 | 1.098 | 0.720 | 0.947 | 0.277 | 0.444 | 0.462 | 0.651 | 0.499 | 0.692 |
| 105 | 259 | 154 | 9.62 | 0.832 | -0.292 | 0.374 | 0.623 | 0.248 | 0.474 | -0.188 | -0.020 | -0.005 | 0.183 | 0.029 | 0.222 |
| 105 | 270 | 165 | 8.02 | 0.745 | 3.556 | 4.707 | 4.975 | 4.558 | 4.799 | 4.119 | 4.299 | 4.301 | 4.503 | 4.316 | 4.520 |
| 105 | 263 | 158 | 8.83 | 0.792 | 1.798 | 2.337 | 2.596 | 2.201 | 2.434 | 1.756 | 1.928 | 1.941 | 2.136 | 1.970 | 2.168 |
| 106 | 259 | 153 | 9.8 | 0.840 | -0.507 | 0.344 | 0.600 | 0.217 | 0.449 | -0.238 | -0.067 | -0.048 | 0.145 | -0.008 | 0.190 |
| 106 | 261 | 155 | 9.71 | 0.833 | -0.728 | 0.335 | 0.591 | 0.206 | 0.438 | -0.245 | -0.072 | -0.056 | 0.138 | -0.020 | 0.179 |
| 106 | 263 | 157 | 9.4 | 0.816 | 0.033 | 1.003 | 1.263 | 0.870 | 1.106 | 0.417 | 0.592 | 0.606 | 0.803 | 0.640 | 0.841 |
| 106 | 267 | 161 | 8.32 | 0.763 | 2.803 | 4.174 | 4.447 | 4.026 | 4.271 | 3.554 | 3.735 | 3.750 | 3.954 | 3.778 | 3.986 |
| 106 | 269 | 163 | 8.7 | 0.777 | 2.681 | 2.708 | 2.978 | 2.565 | 2.808 | 2.115 | 2.296 | 2.302 | 2.506 | 2.325 | 2.532 |
| 106 | 271 | 165 | 8.66 | 0.772 | 2.212 | 2.734 | 3.005 | 2.590 | 2.833 | 2.147 | 2.331 | 2.331 | 2.537 | 2.350 | 2.558 |
| 107 | 260 | 153 | 10.4 | 0.863 | -1.456 | -0.920 | -0.662 | -1.045 | -0.810 | -1.501 | -1.327 | -1.310 | -1.114 | -1.268 | -1.067 |
| 107 | 261 | 154 | 10.5 | 0.865 | -1.928 | -1.316 | -1.060 | -1.440 | -1.206 | -1.890 | -1.715 | -1.701 | -1.505 | -1.661 | -1.460 |
| 107 | 264 | 157 | 9.96 | 0.838 | -0.357 | -0.271 | -0.008 | -0.402 | -0.163 | -0.859 | -0.681 | -0.668 | -0.467 | -0.632 | -0.426 |
| 107 | 265 | 158 | 9.38 | 0.812 | -0.027 | 1.290 | 1.560 | 1.152 | 1.396 | 0.680 | 0.861 | 0.877 | 1.081 | 0.913 | 1.122 |
| 107 | 266 | 159 | 9.43 | 0.813 | 0.398 | 1.036 | 1.306 | 0.898 | 1.143 | 0.432 | 0.614 | 0.627 | 0.831 | 0.661 | 0.870 |
| 107 | 267 | 160 | 8.96 | 0.791 | 1.342 | 2.408 | 2.684 | 2.264 | 2.513 | 1.786 | 1.970 | 1.985 | 2.192 | 2.018 | 2.230 |
| 107 | 270 | 163 | 9.06 | 0.791 | 1.778 | 1.845 | 2.120 | 1.702 | 1.950 | 1.240 | 1.426 | 1.432 | 1.641 | 1.459 | 1.671 |
| 107 | 272 | 165 | 9.14 | 0.792 | 0.913 | 1.470 | 1.744 | 1.326 | 1.575 | 0.876 | 1.063 | 1.063 | 1.273 | 1.085 | 1.298 |
| 107 | 274 | 167 | 8.97 | 0.781 | 1.477 | 1.922 | 2.199 | 1.775 | 2.026 | 1.329 | 1.519 | 1.514 | 1.727 | 1.532 | 1.747 |
| 108 | 264 | 156 | 10.59 | 0.864 | -2.796 | -1.489 | -1.224 | -1.618 | -1.376 | -2.079 | -1.898 | -1.886 | -1.682 | -1.846 | -1.637 |
| 108 | 265 | 157 | 10.47 | 0.857 | -2.708 | -1.309 | -1.042 | -1.439 | -1.196 | -1.901 | -1.719 | -1.707 | -1.503 | -1.669 | -1.459 |
| 108 | 267 | 159 | 10.037 | 0.837 | -1.187 | -0.386 | -0.114 | -0.522 | -0.274 | -0.991 | -0.806 | -0.795 | -0.586 | -0.758 | -0.545 |
| 108 | 268 | 160 | 9.62 | 0.818 | 0.152 | 0.696 | 0.974 | 0.556 | 0.808 | 0.076 | 0.264 | 0.276 | 0.488 | 0.312 | 0.528 |
| 108 | 269 | 161 | 9.32 | 0.804 | 1.182 | 1.506 | 1.788 | 1.361 | 1.616 | 0.875 | 1.065 | 1.078 | 1.291 | 1.112 | 1.330 |
| 108 | 270 | 162 | 9.15 | 0.795 | 0.881 | 1.952 | 2.236 | 1.805 | 2.062 | 1.317 | 1.508 | 1.521 | 1.735 | 1.553 | 1.772 |
| 108 | 273 | 165 | 9.73 | 0.815 | -0.041 | -0.018 | 0.260 | -0.160 | 0.093 | -0.615 | -0.423 | -0.425 | -0.210 | -0.399 | -0.182 |
| 108 | 275 | 167 | 9.44 | 0.800 | -0.538 | 0.735 | 1.017 | 0.589 | 0.845 | 0.134 | 0.328 | 0.323 | 0.541 | 0.345 | 0.565 |
| 109 | 268 | 159 | 10.67 | 0.860 | -1.678 | -1.746 | -1.470 | -1.879 | -1.627 | -2.350 | -2.161 | -2.153 | -1.940 | -2.114 | -1.897 |
| 109 | 274 | 165 | 10.04 | 0.826 | -0.353 | -0.614 | -0.328 | -0.757 | -0.497 | -1.226 | -1.029 | -1.030 | -0.810 | -1.001 | -0.777 |
| 109 | 275 | 166 | 10.48 | 0.842 | -2.013 | -1.857 | -1.577 | -1.995 | -1.739 | -2.446 | -2.250 | -2.257 | -2.039 | -2.231 | -2.009 |
| 109 | 276 | 167 | 9.81 | 0.813 | -0.143 | -0.080 | 0.210 | -0.226 | 0.037 | -0.695 | -0.495 | -0.499 | -0.276 | -0.474 | -0.248 |
| 109 | 278 | 169 | 9.59 | 0.802 | 0.556 | 0.480 | 0.773 | 0.330 | 0.596 | -0.137 | 0.065 | 0.057 | 0.283 | 0.078 | 0.307 |
| 110 | 267 | 157 | 11.78 | 0.905 | -5 | -3.803 | -3.532 | -3.929 | -3.679 | -4.392 | -4.204 | -4.197 | -3.986 | -4.154 | -3.938 |
| 110 | 269 | 159 | 11.51 | 0.891 | -3.747 | -3.458 | -3.182 | -3.587 | -3.334 | -4.053 | -3.862 | -3.857 | -3.642 | -3.817 | -3.597 |
| 110 | 270 | 160 | 11.12 | 0.875 | -3.688 | -2.668 | -2.387 | -2.802 | -2.544 | -3.278 | -3.083 | -3.078 | -2.860 | -3.038 | -2.815 |
| 110 | 271 | 161 | 10.87 | 0.863 | -2.788 | -2.168 | -1.883 | -2.305 | -2.044 | -2.786 | -2.589 | -2.584 | -2.364 | -2.545 | -2.320 |
| 110 | 273 | 163 | 11.37 | 0.879 | -3.77 | -3.551 | -3.272 | -3.683 | -3.427 | -4.140 | -3.945 | -3.948 | -3.730 | -3.914 | -3.692 |
| 110 | 277 | 167 | 10.83 | 0.852 | -2.387 | -2.549 | -2.261 | -2.689 | -2.425 | -3.148 | -2.946 | -2.955 | -2.730 | -2.928 | -2.699 |

| Z | A | N | Q | Temp | Expt | T-IND | T-DEP | T-IND | T-DEP | T-IND | T-DEP | T-IND | T-DEP | T-IND | T-DEP |
|---|---|---|---|---|---|---|---|---|---|---|---|---|---|---|---|
| 110 | 279 | 169 | 9.84 | 0.810 | 0.301 | 0.013 | 0.315 | -0.140 | 0.135 | -0.624 | -0.416 | -0.422 | -0.190 | -0.397 | -0.161 |
| 110 | 281 | 171 | 8.86 | 0.767 | 2.301 | 3.033 | 3.350 | 2.866 | 3.151 | 2.356 | 2.570 | 2.568 | 2.807 | 2.589 | 2.832 |
| 111 | 272 | 161 | 11.197 | 0.874 | -2.42 | -2.678 | -2.385 | -2.817 | -2.548 | -3.311 | -3.109 | -3.103 | -2.877 | -3.061 | -2.829 |
| 111 | 278 | 167 | 10.85 | 0.851 | -2.377 | -2.346 | -2.045 | -2.491 | -2.216 | -2.975 | -2.766 | -2.773 | -2.539 | -2.741 | -2.503 |
| 112 | 281 | 169 | 10.46 | 0.832 | -1 | -1.191 | -0.868 | -1.347 | -1.054 | -1.865 | -1.644 | -1.649 | -1.401 | -1.616 | -1.364 |
| 112 | 283 | 171 | 9.725 | 0.800 | 0.58 | 0.815 | 1.150 | 0.647 | 0.949 | 0.111 | 0.337 | 0.334 | 0.588 | 0.364 | 0.622 |
| 112 | 285 | 173 | 9.328 | 0.781 | 1.462 | 1.995 | 2.338 | 1.820 | 2.128 | 1.277 | 1.508 | 1.502 | 1.761 | 1.528 | 1.790 |
| 113 | 283 | 170 | 10.313 | 0.823 | -1 | -0.576 | -0.234 | -0.742 | -0.432 | -1.290 | -1.059 | -1.061 | -0.802 | -1.026 | -0.761 |
| 113 | 284 | 171 | 10.191 | 0.817 | -0.319 | -0.281 | 0.065 | -0.449 | -0.137 | -0.998 | -0.765 | -0.769 | -0.508 | -0.735 | -0.469 |
| 113 | 285 | 172 | 9.927 | 0.805 | 0.738 | 0.445 | 0.795 | 0.272 | 0.588 | -0.283 | -0.048 | -0.052 | 0.212 | -0.020 | 0.249 |
| 114 | 286 | 172 | 10.373 | 0.821 | -0.886 | -0.572 | -0.210 | -0.746 | -0.419 | -1.315 | -1.072 | -1.078 | -0.805 | -1.042 | -0.764 |
| 114 | 287 | 173 | 10.211 | 0.813 | -0.319 | -0.152 | 0.214 | -0.329 | 0.000 | -0.900 | -0.656 | -0.663 | -0.388 | -0.629 | -0.348 |
| 114 | 288 | 174 | 10.08 | 0.807 | -0.097 | 0.197 | 0.567 | 0.017 | 0.349 | -0.556 | -0.309 | -0.318 | -0.040 | -0.286 | -0.004 |
| 114 | 289 | 175 | 10.007 | 0.802 | 0.415 | 0.389 | 0.762 | 0.207 | 0.541 | -0.364 | -0.116 | -0.127 | 0.152 | -0.097 | 0.186 |
| 115 | 287 | 172 | 10.789 | 0.836 | -1.456 | -1.444 | -1.069 | -1.620 | -1.282 | -2.205 | -1.954 | -1.961 | -1.679 | -1.921 | -1.633 |
| 115 | 288 | 173 | 10.677 | 0.830 | -1.06 | -1.188 | -0.810 | -1.367 | -1.026 | -1.952 | -1.700 | -1.709 | -1.424 | -1.670 | -1.380 |
| 115 | 289 | 174 | 10.504 | 0.822 | -0.658 | -0.751 | -0.368 | -0.933 | -0.589 | -1.522 | -1.267 | -1.278 | -0.990 | -1.241 | -0.949 |
| 116 | 290 | 174 | 11.042 | 0.841 | -2.167 | -1.940 | -1.544 | -2.123 | -1.766 | -2.724 | -2.461 | -2.474 | -2.177 | -2.433 | -2.131 |
| 116 | 291 | 175 | 10.94 | 0.835 | -1.699 | -1.709 | -1.308 | -1.893 | -1.534 | -2.495 | -2.230 | -2.245 | -1.946 | -2.207 | -1.902 |
| 116 | 292 | 176 | 10.858 | 0.831 | -1.397 | -1.518 | -1.114 | -1.705 | -1.343 | -2.307 | -2.039 | -2.057 | -1.755 | -2.021 | -1.714 |
| 116 | 293 | 177 | 10.736 | 0.825 | -1.097 | -1.209 | -0.800 | -1.399 | -1.033 | -2.002 | -1.732 | -1.752 | -1.448 | -1.718 | -1.408 |
| 117 | 293 | 176 | 11.233 | 0.843 | -1.824 | -2.252 | -1.829 | -2.443 | -2.064 | -3.065 | -2.787 | -2.806 | -2.493 | -2.766 | -2.446 |

| Parent Nuclei | | | $Q_\alpha(Mev)$ | Temp (Mev) | $log_{10}T_{1/2}/s$ | | | | | | | | | | |
|---|---|---|---|---|---|---|---|---|---|---|---|---|---|---|---|
| | | | | | Expt | Prox.77-6 | | Prox.77-7 | | Prox.77-8 | | Prox.77-9 | | Prox.77-10 | |
| Z | A | N | | | | T-IND | T-DEP | T-IND | T-DEP | T-IND | T-DEP | T-IND | T-DEP | T-IND | T-DEP |
| 104 | 259 | 155 | 9.13 | 0.811 | 0.415 | 0.903 | 1.080 | 0.924 | 1.103 | 1.425 | 1.687 | 0.919 | 1.097 | 0.918 | 1.096 |
| 104 | 263 | 159 | 8.25 | 0.766 | 3.301 | 3.400 | 3.580 | 3.421 | 3.605 | 3.970 | 4.248 | 3.414 | 3.597 | 3.414 | 3.596 |
| 105 | 256 | 151 | 9.34 | 0.825 | 0.453 | 1.135 | 1.315 | 1.158 | 1.341 | 1.656 | 1.921 | 1.153 | 1.336 | 1.152 | 1.334 |
| 105 | 258 | 153 | 9.5 | 0.829 | 0.747 | 0.392 | 0.572 | 0.414 | 0.597 | 0.909 | 1.173 | 0.409 | 0.591 | 0.408 | 0.590 |
| 105 | 259 | 154 | 9.62 | 0.832 | -0.292 | -0.076 | 0.104 | -0.054 | 0.128 | 0.438 | 0.700 | -0.060 | 0.122 | -0.061 | 0.121 |
| 105 | 270 | 165 | 8.02 | 0.745 | 3.556 | 4.203 | 4.392 | 4.225 | 4.417 | 4.817 | 5.111 | 4.216 | 4.407 | 4.216 | 4.407 |
| 105 | 263 | 158 | 8.83 | 0.792 | 1.798 | 1.861 | 2.045 | 1.883 | 2.070 | 2.417 | 2.693 | 1.876 | 2.062 | 1.875 | 2.061 |
| 106 | 259 | 153 | 9.8 | 0.840 | -0.507 | -0.117 | 0.068 | -0.094 | 0.093 | 0.404 | 0.671 | -0.099 | 0.088 | -0.101 | 0.086 |
| 106 | 261 | 155 | 9.71 | 0.833 | -0.728 | -0.129 | 0.057 | -0.106 | 0.082 | 0.399 | 0.669 | -0.111 | 0.076 | -0.113 | 0.075 |
| 106 | 263 | 157 | 9.4 | 0.816 | 0.033 | 0.530 | 0.717 | 0.552 | 0.743 | 1.075 | 1.350 | 0.546 | 0.736 | 0.545 | 0.735 |
| 106 | 267 | 161 | 8.32 | 0.763 | 2.803 | 3.661 | 3.854 | 3.684 | 3.880 | 4.266 | 4.560 | 3.677 | 3.872 | 3.676 | 3.871 |
| 106 | 269 | 163 | 8.7 | 0.777 | 2.681 | 2.211 | 2.404 | 2.234 | 2.430 | 2.803 | 3.094 | 2.226 | 2.421 | 2.226 | 2.420 |
| 106 | 271 | 165 | 8.66 | 0.772 | 2.212 | 2.237 | 2.431 | 2.259 | 2.456 | 2.836 | 3.129 | 2.251 | 2.446 | 2.250 | 2.446 |
| 107 | 260 | 153 | 10.4 | 0.863 | -1.456 | -1.377 | -1.189 | -1.354 | -1.163 | -0.865 | -0.597 | -1.359 | -1.168 | -1.360 | -1.170 |
| 107 | 261 | 154 | 10.5 | 0.865 | -1.928 | -1.769 | -1.581 | -1.746 | -1.556 | -1.259 | -0.992 | -1.751 | -1.561 | -1.753 | -1.563 |
| 107 | 264 | 157 | 9.96 | 0.838 | -0.357 | -0.742 | -0.550 | -0.719 | -0.525 | -0.204 | 0.072 | -0.725 | -0.531 | -0.726 | -0.533 |
| 107 | 265 | 158 | 9.38 | 0.812 | -0.027 | 0.798 | 0.993 | 0.822 | 1.019 | 1.363 | 1.649 | 0.816 | 1.012 | 0.814 | 1.011 |
| 107 | 266 | 159 | 9.43 | 0.813 | 0.398 | 0.547 | 0.741 | 0.570 | 0.767 | 1.112 | 1.398 | 0.564 | 0.761 | 0.563 | 0.759 |
| 107 | 267 | 160 | 8.96 | 0.791 | 1.342 | 1.901 | 2.098 | 1.925 | 2.124 | 2.490 | 2.783 | 1.918 | 2.117 | 1.917 | 2.116 |
| 107 | 270 | 163 | 9.06 | 0.791 | 1.778 | 1.343 | 1.541 | 1.367 | 1.567 | 1.936 | 2.230 | 1.359 | 1.559 | 1.358 | 1.558 |
| 107 | 272 | 165 | 9.14 | 0.792 | 0.913 | 0.971 | 1.170 | 0.994 | 1.195 | 1.565 | 1.861 | 0.986 | 1.187 | 0.986 | 1.186 |
| 107 | 274 | 167 | 8.97 | 0.781 | 1.477 | 1.418 | 1.618 | 1.440 | 1.643 | 2.026 | 2.326 | 1.432 | 1.634 | 1.431 | 1.633 |
| 108 | 264 | 156 | 10.59 | 0.864 | -2.796 | -1.956 | -1.761 | -1.933 | -1.736 | -1.430 | -1.153 | -1.938 | -1.741 | -1.940 | -1.743 |
| 108 | 265 | 157 | 10.47 | 0.857 | -2.708 | -1.780 | -1.584 | -1.757 | -1.558 | -1.246 | -0.967 | -1.762 | -1.564 | -1.763 | -1.566 |
| 108 | 267 | 159 | 10.037 | 0.837 | -1.187 | -0.872 | -0.673 | -0.849 | -0.647 | -0.316 | -0.029 | -0.854 | -0.653 | -0.856 | -0.655 |
| 108 | 268 | 160 | 9.62 | 0.818 | 0.152 | 0.195 | 0.396 | 0.219 | 0.423 | 0.772 | 1.066 | 0.213 | 0.416 | 0.212 | 0.415 |
| 108 | 269 | 161 | 9.32 | 0.804 | 1.182 | 0.994 | 1.196 | 1.018 | 1.224 | 1.587 | 1.886 | 1.011 | 1.216 | 1.010 | 1.215 |
| 108 | 270 | 162 | 9.15 | 0.795 | 0.881 | 1.434 | 1.638 | 1.458 | 1.665 | 2.037 | 2.340 | 1.451 | 1.657 | 1.450 | 1.656 |
| 108 | 273 | 165 | 9.73 | 0.815 | -0.041 | -0.513 | -0.310 | -0.491 | -0.285 | 0.071 | 0.368 | -0.498 | -0.293 | -0.499 | -0.294 |
| 108 | 275 | 167 | 9.44 | 0.800 | -0.538 | 0.229 | 0.435 | 0.252 | 0.461 | 0.834 | 1.137 | 0.244 | 0.451 | 0.244 | 0.451 |
| 109 | 268 | 159 | 10.67 | 0.860 | -1.678 | -2.227 | -2.024 | -2.204 | -1.998 | -1.680 | -1.392 | -2.209 | -2.004 | -2.211 | -2.006 |
| 109 | 274 | 165 | 10.04 | 0.826 | -0.353 | -1.117 | -0.908 | -1.094 | -0.882 | -0.528 | -0.224 | -1.101 | -0.890 | -1.102 | -0.891 |
| 109 | 275 | 166 | 10.48 | 0.842 | -2.013 | -2.344 | -2.136 | -2.321 | -2.111 | -1.771 | -1.474 | -2.328 | -2.119 | -2.329 | -2.120 |
| 109 | 276 | 167 | 9.81 | 0.813 | -0.143 | -0.592 | -0.380 | -0.568 | -0.354 | 0.014 | 0.324 | -0.576 | -0.363 | -0.577 | -0.364 |
| 109 | 278 | 169 | 9.59 | 0.802 | 0.556 | -0.040 | 0.173 | -0.017 | 0.200 | 0.583 | 0.898 | -0.025 | 0.190 | -0.026 | 0.190 |
| 110 | 267 | 157 | 11.78 | 0.905 | -5 | -4.265 | -4.062 | -4.242 | -4.036 | -3.749 | -3.467 | -4.246 | -4.041 | -4.248 | -4.043 |
| 110 | 269 | 159 | 11.51 | 0.891 | -3.747 | -3.928 | -3.723 | -3.905 | -3.697 | -3.398 | -3.111 | -3.910 | -3.702 | -3.912 | -3.704 |
| 110 | 270 | 160 | 11.12 | 0.875 | -3.688 | -3.153 | -2.944 | -3.129 | -2.917 | -2.604 | -2.310 | -3.134 | -2.923 | -3.136 | -2.925 |
| 110 | 271 | 161 | 10.87 | 0.863 | -2.788 | -2.661 | -2.451 | -2.637 | -2.424 | -2.099 | -1.801 | -2.643 | -2.430 | -2.644 | -2.432 |
| 110 | 273 | 163 | 11.37 | 0.879 | -3.77 | -4.026 | -3.817 | -4.003 | -3.792 | -3.481 | -3.188 | -4.009 | -3.798 | -4.010 | -3.800 |
| 110 | 277 | 167 | 10.83 | 0.852 | -2.387 | -3.043 | -2.828 | -3.019 | -2.803 | -2.464 | -2.158 | -3.027 | -2.811 | -3.027 | -2.812 |
| 110 | 279 | 169 | 9.84 | 0.810 | 0.301 | -0.519 | -0.299 | -0.495 | -0.271 | 0.113 | 0.437 | -0.503 | -0.281 | -0.504 | -0.282 |
| 110 | 281 | 171 | 8.86 | 0.767 | 2.301 | 2.460 | 2.685 | 2.485 | 2.714 | 3.150 | 3.495 | 2.476 | 2.704 | 2.475 | 2.703 |
| 111 | 272 | 161 | 11.197 | 0.874 | -2.42 | -3.180 | -2.963 | -3.155 | -2.935 | -2.612 | -2.306 | -3.161 | -2.941 | -3.162 | -2.943 |
| 111 | 278 | 167 | 10.85 | 0.851 | -2.377 | -2.861 | -2.638 | -2.836 | -2.611 | -2.262 | -1.943 | -2.843 | -2.619 | -2.844 | -2.620 |
| 112 | 281 | 169 | 10.46 | 0.832 | -1 | -1.744 | -1.509 | -1.718 | -1.479 | -1.097 | -0.754 | -1.725 | -1.488 | -1.727 | -1.489 |
| 112 | 283 | 171 | 9.725 | 0.800 | 0.58 | 0.230 | 0.470 | 0.257 | 0.500 | 0.922 | 1.282 | 0.248 | 0.491 | 0.247 | 0.490 |
| 112 | 285 | 173 | 9.328 | 0.781 | 1.462 | 1.391 | 1.635 | 1.419 | 1.666 | 2.114 | 2.486 | 1.409 | 1.655 | 1.408 | 1.654 |
| 113 | 283 | 170 | 10.313 | 0.823 | -1 | -1.161 | -0.914 | -1.133 | -0.883 | -0.478 | -0.114 | -1.141 | -0.892 | -1.142 | -0.894 |
| 113 | 284 | 171 | 10.191 | 0.817 | -0.319 | -0.871 | -0.623 | -0.844 | -0.592 | -0.178 | 0.191 | -0.852 | -0.602 | -0.853 | -0.603 |
| 113 | 285 | 172 | 9.927 | 0.805 | 0.738 | -0.158 | 0.092 | -0.130 | 0.124 | 0.553 | 0.930 | -0.139 | 0.114 | -0.140 | 0.112 |
| 114 | 286 | 172 | 10.373 | 0.821 | -0.886 | -1.182 | -0.923 | -1.154 | -0.891 | -0.468 | -0.081 | -1.162 | -0.901 | -1.164 | -0.902 |
| 114 | 287 | 173 | 10.211 | 0.813 | -0.319 | -0.770 | -0.510 | -0.742 | -0.477 | -0.042 | 0.351 | -0.751 | -0.487 | -0.752 | -0.489 |
| 114 | 288 | 174 | 10.08 | 0.807 | -0.097 | -0.429 | -0.166 | -0.400 | -0.134 | 0.311 | 0.710 | -0.409 | -0.144 | -0.410 | -0.145 |
| 114 | 289 | 175 | 10.007 | 0.802 | 0.415 | -0.241 | 0.023 | -0.212 | 0.056 | 0.508 | 0.912 | -0.222 | 0.045 | -0.223 | 0.044 |
| 115 | 287 | 172 | 10.789 | 0.836 | -1.456 | -2.064 | -1.796 | -2.035 | -1.763 | -1.344 | -0.944 | -2.043 | -1.772 | -2.045 | -1.774 |

| Z | A | N | $Q_\alpha$(Mev) | Temp (Mev) | | | | | | | | | | |
|---|---|---|---|---|---|---|---|---|---|---|---|---|---|---|
| 115 | 288 | 173 | 10.677 | 0.830 | -1.06 | -1.815 | -1.545 | -1.785 | -1.512 | -1.083 | -0.679 | -1.794 | -1.521 | -1.795 | -1.523 |
| 115 | 289 | 174 | 10.504 | 0.822 | -0.658 | -1.387 | -1.115 | -1.357 | -1.081 | -0.641 | -0.229 | -1.366 | -1.092 | -1.368 | -1.093 |
| 116 | 290 | 174 | 11.042 | 0.841 | -2.167 | -2.581 | -2.299 | -2.551 | -2.265 | -1.835 | -1.411 | -2.560 | -2.275 | -2.561 | -2.277 |
| 116 | 291 | 175 | 10.94 | 0.835 | -1.699 | -2.355 | -2.072 | -2.325 | -2.038 | -1.598 | -1.168 | -2.334 | 2.048 | -2.335 | 2.049 |
| 116 | 292 | 176 | 10.858 | 0.831 | -1.397 | -2.170 | -1.885 | -2.140 | -1.850 | -1.403 | -0.967 | -2.149 | -1.861 | -2.150 | -1.862 |
| 116 | 293 | 177 | 10.736 | 0.825 | -1.097 | -1.868 | -1.581 | -1.838 | -1.546 | -1.088 | -0.646 | -1.848 | -1.557 | -1.849 | -1.559 |
| 117 | 293 | 176 | 11.233 | 0.843 | -1.824 | -2.919 | -2.622 | -2.888 | -2.586 | -2.139 | -1.685 | -2.897 | -2.597 | -2.898 | -2.598 |

| Parent Nuclei | | | $Q_\alpha$(Mev) | Temp (Mev) | $log_{10}T_{1/2}/s$ | | | | | | | | | |
|---|---|---|---|---|---|---|---|---|---|---|---|---|---|---|
| | | | | | Expt | Prox.77-11 | | Prox.77-12 | | Prox.77-13 | | Prox.81 | | Bass73 | |
| Z | A | N | | | | T-IND | T-DEP | T-IND | T-DEP | T-IND | T-DEP | T-IND | T-DEP | T-IND | T-DEP |
| 104 | 259 | 155 | 9.13 | 0.811 | 0.415 | 1.119 | 1.324 | 1.342 | 1.586 | 1.490 | 1.768 | 1.367 | 1.622 | 1.260 | 1.255 |
| 104 | 263 | 159 | 8.25 | 0.766 | 3.301 | 3.627 | 3.838 | 3.855 | 4.107 | 4.037 | 4.335 | 3.898 | 4.161 | 3.793 | 3.789 |
| 105 | 256 | 151 | 9.34 | 0.825 | 0.453 | 1.358 | 1.569 | 1.600 | 1.853 | 1.725 | 2.008 | 1.610 | 1.871 | 1.531 | 1.526 |
| 105 | 258 | 153 | 9.5 | 0.829 | 0.747 | 0.611 | 0.821 | 0.845 | 1.095 | 0.977 | 1.256 | 0.859 | 1.118 | 0.777 | 0.772 |
| 105 | 259 | 154 | 9.62 | 0.832 | -0.292 | 0.141 | 0.350 | 0.369 | 0.617 | 0.504 | 0.782 | 0.385 | 0.644 | 0.301 | 0.296 |
| 105 | 270 | 165 | 8.02 | 0.745 | 3.556 | 4.439 | 4.660 | 4.665 | 4.925 | 4.884 | 5.198 | 4.725 | 5.001 | 4.630 | 4.626 |
| 105 | 263 | 158 | 8.83 | 0.792 | 1.798 | 2.088 | 2.302 | 2.320 | 2.575 | 2.485 | 2.778 | 2.352 | 2.619 | 2.271 | 2.266 |
| 106 | 259 | 153 | 9.8 | 0.840 | -0.507 | 0.106 | 0.321 | 0.346 | 0.601 | 0.472 | 0.756 | 0.355 | 0.620 | 0.296 | 0.290 |
| 106 | 261 | 155 | 9.71 | 0.833 | -0.728 | 0.095 | 0.310 | 0.331 | 0.587 | 0.467 | 0.753 | 0.345 | 0.612 | 0.285 | 0.280 |
| 106 | 263 | 157 | 9.4 | 0.816 | 0.033 | 0.757 | 0.975 | 0.994 | 1.252 | 1.143 | 1.435 | 1.016 | 1.285 | 0.955 | 0.950 |
| 106 | 267 | 161 | 8.32 | 0.763 | 2.803 | 3.904 | 4.130 | 4.151 | 4.419 | 4.338 | 4.653 | 4.192 | 4.472 | 4.126 | 4.122 |
| 106 | 269 | 163 | 8.7 | 0.777 | 2.681 | 2.447 | 2.671 | 2.680 | 2.943 | 2.871 | 3.179 | 2.724 | 3.002 | 2.654 | 2.649 |
| 106 | 271 | 165 | 8.66 | 0.772 | 2.212 | 2.472 | 2.697 | 2.699 | 2.962 | 2.903 | 3.213 | 2.750 | 3.029 | 2.674 | 2.669 |
| 107 | 260 | 153 | 10.4 | 0.863 | -1.456 | -1.155 | -0.937 | -0.915 | -0.656 | -0.797 | -0.514 | -0.913 | -0.643 | -0.955 | -0.961 |
| 107 | 261 | 154 | 10.5 | 0.865 | -1.928 | -1.549 | -1.332 | -1.313 | -1.057 | -1.192 | -0.910 | -1.310 | -1.041 | -1.355 | -1.360 |
| 107 | 264 | 157 | 9.96 | 0.838 | -0.357 | -0.515 | -0.293 | -0.276 | -0.015 | -0.136 | 0.156 | -0.262 | 0.013 | -0.306 | -0.311 |
| 107 | 265 | 158 | 9.38 | 0.812 | -0.027 | 1.034 | 1.260 | 1.282 | 1.550 | 1.435 | 1.737 | 1.303 | 1.583 | 1.261 | 1.256 |
| 107 | 266 | 159 | 9.43 | 0.813 | 0.398 | 0.782 | 1.008 | 1.025 | 1.292 | 1.182 | 1.484 | 1.049 | 1.329 | 1.004 | 0.999 |
| 107 | 267 | 160 | 8.96 | 0.791 | 1.342 | 2.143 | 2.372 | 2.393 | 2.666 | 2.563 | 2.874 | 2.423 | 2.708 | 2.377 | 2.372 |
| 107 | 270 | 163 | 9.06 | 0.791 | 1.778 | 1.582 | 1.812 | 1.821 | 2.092 | 2.005 | 2.317 | 1.860 | 2.144 | 1.805 | 1.800 |
| 107 | 272 | 165 | 9.14 | 0.792 | 0.913 | 1.208 | 1.438 | 1.439 | 1.708 | 1.633 | 1.945 | 1.484 | 1.768 | 1.422 | 1.417 |
| 107 | 274 | 167 | 8.97 | 0.781 | 1.477 | 1.656 | 1.888 | 1.884 | 2.154 | 2.093 | 2.409 | 1.937 | 2.224 | 1.868 | 1.863 |
| 108 | 264 | 156 | 10.59 | 0.864 | -2.796 | -1.730 | -1.505 | -1.488 | -1.223 | -1.361 | -1.070 | -1.484 | -1.205 | -1.513 | -1.518 |
| 108 | 265 | 157 | 10.47 | 0.857 | -2.708 | -1.552 | -1.326 | -1.310 | -1.044 | -1.178 | -0.884 | -1.303 | -1.023 | -1.332 | -1.338 |
| 108 | 267 | 159 | 10.037 | 0.837 | -1.187 | -0.638 | -0.408 | -0.393 | -0.122 | -0.246 | 0.057 | -0.377 | -0.092 | -0.408 | -0.414 |
| 108 | 268 | 160 | 9.62 | 0.818 | 0.152 | 0.436 | 0.669 | 0.686 | 0.962 | 0.844 | 1.155 | 0.709 | 0.997 | 0.676 | 0.670 |
| 108 | 269 | 161 | 9.32 | 0.804 | 1.182 | 1.239 | 1.475 | 1.493 | 1.772 | 1.660 | 1.977 | 1.520 | 1.812 | 1.484 | 1.479 |
| 108 | 270 | 162 | 9.15 | 0.795 | 0.881 | 1.681 | 1.918 | 1.936 | 2.216 | 2.111 | 2.432 | 1.967 | 2.261 | 1.928 | 1.923 |
| 108 | 273 | 165 | 9.73 | 0.815 | -0.041 | -0.277 | -0.043 | -0.043 | 0.230 | 0.139 | 0.451 | -0.007 | 0.283 | -0.057 | -0.062 |
| 108 | 275 | 167 | 9.44 | 0.800 | -0.538 | 0.469 | 0.707 | 0.703 | 0.979 | 0.901 | 1.221 | 0.748 | 1.041 | 0.690 | 0.685 |

| Z | A | N | $Q_\alpha$(Mev) | Temp (Mev) | Expt | Bass77 T-IND | Bass77 T-DEP | Bass80 T-IND | Bass80 T-DEP | CW76 T-IND | CW76 T-DEP | BW91 T-IND | BW91 T-DEP | AW95 T-IND | AW95 T-DEP |
|---|---|---|---|---|---|---|---|---|---|---|---|---|---|---|---|
| 109 | 268 | 159 | 10.67 | 0.860 | -1.678 | -1.994 | -1.760 | -1.749 | -1.474 | -1.610 | -1.307 | -1.740 | -1.451 | -1.762 | -1.767 |
| 109 | 274 | 165 | 10.04 | 0.826 | -0.353 | -0.876 | -0.635 | -0.635 | -0.353 | -0.458 | -0.139 | -0.604 | -0.306 | -0.645 | -0.650 |
| 109 | 275 | 166 | 10.48 | 0.842 | -2.013 | -2.111 | -1.872 | -1.880 | -1.604 | -1.705 | -1.393 | -1.850 | -1.556 | -1.895 | -1.900 |
| 109 | 276 | 167 | 9.81 | 0.813 | -0.143 | -0.348 | -0.104 | -0.107 | 0.177 | 0.084 | 0.410 | -0.068 | 0.233 | -0.118 | -0.124 |
| 109 | 278 | 169 | 9.59 | 0.802 | 0.556 | 0.206 | 0.453 | 0.445 | 0.732 | 0.652 | 0.984 | 0.493 | 0.798 | 0.432 | 0.427 |
| 110 | 267 | 157 | 11.78 | 0.905 | -5 | -4.039 | -3.806 | -3.797 | -3.525 | -3.681 | -3.386 | -3.806 | -3.516 | -3.819 | -3.825 |
| 110 | 269 | 159 | 11.51 | 0.891 | -3.747 | -3.699 | -3.463 | -3.457 | -3.181 | -3.329 | -3.028 | -3.458 | -3.165 | -3.474 | -3.480 |
| 110 | 270 | 160 | 11.12 | 0.875 | -3.688 | -2.918 | -2.678 | -2.670 | -2.389 | -2.533 | -2.225 | -2.665 | -2.368 | -2.682 | -2.688 |
| 110 | 271 | 161 | 10.87 | 0.863 | -2.788 | -2.423 | -2.180 | -2.172 | -1.888 | -2.028 | -1.714 | -2.163 | -1.863 | -2.182 | -2.188 |
| 110 | 273 | 163 | 11.37 | 0.879 | -3.77 | -3.796 | -3.558 | -3.561 | -3.284 | -3.414 | -3.107 | -3.551 | -3.255 | -3.579 | -3.585 |
| 110 | 277 | 167 | 10.83 | 0.852 | -2.387 | -2.806 | -2.560 | -2.571 | -2.286 | -2.396 | -2.076 | -2.544 | -2.240 | -2.587 | -2.593 |
| 110 | 279 | 169 | 9.84 | 0.810 | 0.301 | -0.266 | -0.012 | -0.018 | 0.279 | 0.185 | 0.526 | 0.026 | 0.340 | -0.033 | -0.039 |
| 110 | 281 | 171 | 8.86 | 0.767 | 2.301 | 2.730 | 2.993 | 2.993 | 3.301 | 3.227 | 3.594 | 3.052 | 3.379 | 2.968 | 2.963 |
| 111 | 272 | 161 | 11.197 | 0.874 | -2.42 | -2.936 | -2.687 | -2.679 | -2.386 | -2.539 | -2.218 | -2.675 | -2.365 | -2.689 | -2.695 |
| 111 | 278 | 167 | 10.85 | 0.851 | -2.377 | -2.614 | -2.358 | -2.364 | -2.067 | 2.190 | -1.856 | -2.340 | -2.023 | -2.383 | -2.389 |
| 112 | 281 | 169 | 10.46 | 0.832 | -1 | -1.479 | -1.207 | -1.212 | -0.893 | -1.020 | -0.659 | -1.179 | -0.843 | -1.245 | -1.250 |
| 112 | 283 | 171 | 9.725 | 0.800 | 0.58 | 0.508 | 0.787 | 0.785 | 1.115 | 1.003 | 1.385 | 0.832 | 1.179 | 0.740 | 0.734 |
| 112 | 285 | 173 | 9.328 | 0.781 | 1.462 | 1.676 | 1.960 | 1.957 | 2.292 | 2.196 | 2.593 | 2.015 | 2.369 | 1.897 | 1.892 |
| 113 | 283 | 170 | 10.313 | 0.823 | -1 | -0.881 | -0.595 | -0.596 | -0.258 | -0.395 | -0.010 | -0.561 | -0.206 | -0.652 | -0.658 |
| 113 | 284 | 171 | 10.191 | 0.817 | -0.319 | -0.590 | -0.302 | -0.305 | 0.035 | -0.095 | 0.295 | -0.265 | 0.093 | -0.367 | -0.372 |
| 113 | 285 | 172 | 9.927 | 0.805 | 0.738 | 0.128 | 0.420 | 0.416 | 0.761 | 0.638 | 1.037 | 0.463 | 0.825 | 0.346 | 0.341 |
| 114 | 286 | 172 | 10.373 | 0.821 | -0.886 | -0.891 | -0.590 | -0.595 | -0.238 | -0.381 | 0.029 | -0.555 | -0.180 | -0.687 | -0.693 |
| 114 | 287 | 173 | 10.211 | 0.813 | -0.319 | -0.476 | -0.173 | -0.179 | 0.181 | 0.045 | 0.462 | -0.134 | 0.245 | -0.280 | -0.285 |
| 114 | 288 | 174 | 10.08 | 0.807 | -0.097 | -0.132 | 0.174 | 0.165 | 0.528 | 0.399 | 0.823 | 0.216 | 0.598 | 0.056 | 0.051 |
| 114 | 289 | 175 | 10.007 | 0.802 | 0.415 | 0.057 | 0.365 | 0.353 | 0.718 | 0.596 | 1.025 | 0.409 | 0.793 | 0.236 | 0.231 |
| 115 | 287 | 172 | 10.789 | 0.836 | -1.456 | -1.768 | -1.456 | -1.462 | -1.091 | -1.255 | -0.831 | -1.428 | -1.039 | -1.581 | -1.587 |
| 115 | 288 | 173 | 10.677 | 0.830 | -1.06 | -1.516 | -1.202 | -1.211 | -0.837 | -0.994 | -0.564 | -1.171 | -0.779 | -1.338 | -1.344 |
| 115 | 289 | 174 | 10.504 | 0.822 | -0.658 | -1.085 | -0.767 | -0.778 | -0.400 | -0.551 | -0.112 | -0.733 | -0.336 | -0.916 | -0.921 |
| 116 | 290 | 174 | 11.042 | 0.841 | -2.167 | -2.276 | -1.947 | -1.961 | -1.570 | -1.743 | -1.292 | -1.922 | -1.512 | -2.130 | -2.136 |
| 116 | 291 | 175 | 10.94 | 0.835 | -1.699 | -2.048 | -1.717 | -1.733 | -1.339 | -1.506 | -1.048 | -1.690 | -1.276 | -1.913 | -1.918 |
| 116 | 292 | 176 | 10.858 | 0.831 | -1.397 | -1.861 | -1.527 | -1.547 | -1.150 | -1.310 | -0.846 | -1.498 | -1.081 | -1.737 | -1.742 |
| 116 | 293 | 177 | 10.736 | 0.825 | -1.097 | -1.556 | -1.220 | -1.242 | -0.841 | -0.995 | -0.523 | -1.188 | -0.767 | -1.444 | -1.449 |
| 117 | 293 | 176 | 11.233 | 0.843 | -1.824 | -2.602 | -2.254 | -2.275 | -1.859 | -2.043 | -1.557 | -2.231 | -1.795 | -2.508 | -2.514 |

| Parent Nuclei | | | $Q_\alpha$(Mev) | Temp (Mev) | $log_{10}T_{1/2}/s$ | | | | | | | | | |
|---|---|---|---|---|---|---|---|---|---|---|---|---|---|---|
| | | | | | Expt | Bass77 | | Bass80 | | CW76 | | BW91 | | AW95 | |
| Z | A | N | | | | T-IND | T-DEP | T-IND | T-DEP | T-IND | T-DEP | T-IND | T-DEP | T-IND | T-DEP |
| 104 | 259 | 155 | 9.13 | 0.811 | 0.415 | 2.194 | 2.187 | 0.890 | 0.883 | 0.280 | 0.275 | 0.879 | 1.114 | -0.131 | 0.049 |
| 104 | 263 | 159 | 8.25 | 0.766 | 3.301 | 4.733 | 4.727 | 3.394 | 3.388 | 2.733 | 2.728 | 3.377 | 3.612 | 2.340 | 2.518 |
| 105 | 256 | 151 | 9.34 | 0.825 | 0.453 | 2.475 | 2.468 | 1.163 | 1.157 | 0.538 | 0.532 | 1.148 | 1.391 | 0.126 | 0.312 |
| 105 | 258 | 153 | 9.5 | 0.829 | 0.747 | 1.712 | 1.704 | 0.407 | 0.401 | -0.200 | -0.206 | 0.396 | 0.635 | -0.614 | -0.431 |

| | | | | | | | | | | | | | | |
|---|---|---|---|---|---|---|---|---|---|---|---|---|---|---|
| 105 | 259 | 154 | 9.62 | 0.832 | -0.292 | 1.230 | 1.223 | -0.068 | -0.075 | -0.664 | -0.670 | -0.077 | 0.160 | -1.080 | -0.898 |
| 105 | 270 | 165 | 8.02 | 0.745 | 3.556 | 5.560 | 5.554 | 4.199 | 4.193 | 3.521 | 3.516 | 4.182 | 4.415 | 3.138 | 3.314 |
| 105 | 263 | 158 | 8.83 | 0.792 | 1.798 | 3.204 | 3.197 | 1.874 | 1.868 | 1.234 | 1.228 | 1.859 | 2.097 | 0.833 | 1.014 |
| 106 | 259 | 153 | 9.8 | 0.840 | -0.507 | 1.226 | 1.219 | -0.080 | -0.087 | -0.684 | -0.691 | -0.092 | 0.150 | -1.100 | -0.915 |
| 106 | 261 | 155 | 9.71 | 0.833 | -0.728 | 1.211 | 1.204 | -0.098 | -0.105 | -0.702 | -0.708 | -0.108 | 0.132 | -1.116 | -0.932 |
| 106 | 263 | 157 | 9.4 | 0.816 | 0.033 | 1.881 | 1.874 | 0.560 | 0.553 | -0.060 | -0.066 | 0.548 | 0.787 | -0.467 | -0.285 |
| 106 | 267 | 161 | 8.32 | 0.763 | 2.803 | 5.064 | 5.058 | 3.699 | 3.693 | 3.012 | 3.007 | 3.678 | 3.919 | 2.628 | 2.808 |
| 106 | 269 | 163 | 8.7 | 0.777 | 2.681 | 3.576 | 3.570 | 2.228 | 2.222 | 1.576 | 1.571 | 2.214 | 2.450 | 1.184 | 1.363 |
| 106 | 271 | 165 | 8.66 | 0.772 | 2.212 | 3.591 | 3.585 | 2.242 | 2.236 | 1.594 | 1.588 | 2.229 | 2.464 | 1.202 | 1.380 |
| 107 | 260 | 153 | 10.4 | 0.863 | -1.456 | -0.034 | -0.042 | -1.330 | -1.337 | -1.909 | -1.916 | -1.337 | -1.096 | -2.332 | -2.145 |
| 107 | 261 | 154 | 10.5 | 0.865 | -1.928 | -0.438 | -0.446 | -1.729 | -1.737 | -2.298 | -2.305 | -1.734 | -1.494 | -2.722 | -2.537 |
| 107 | 264 | 157 | 9.96 | 0.838 | -0.357 | 0.611 | 0.604 | -0.701 | -0.708 | -1.299 | -1.305 | -0.709 | -0.470 | -1.713 | -1.528 |
| 107 | 265 | 158 | 9.38 | 0.812 | -0.027 | 2.186 | 2.179 | 0.850 | 0.843 | 0.213 | 0.207 | 0.835 | 1.078 | -0.189 | -0.005 |
| 107 | 266 | 159 | 9.43 | 0.813 | 0.398 | 1.926 | 1.918 | 0.592 | 0.585 | -0.038 | -0.044 | 0.578 | 0.820 | -0.441 | -0.258 |
| 107 | 267 | 160 | 8.96 | 0.791 | 1.342 | 3.306 | 3.299 | 1.952 | 1.946 | 1.291 | 1.286 | 1.934 | 2.178 | 0.898 | 1.081 |
| 107 | 270 | 163 | 9.06 | 0.791 | 1.778 | 2.722 | 2.715 | 1.374 | 1.368 | 0.730 | 0.725 | 1.361 | 1.600 | 0.335 | 0.516 |
| 107 | 272 | 165 | 9.14 | 0.792 | 0.913 | 2.331 | 2.324 | 0.988 | 0.981 | 0.356 | 0.350 | 0.977 | 1.213 | -0.041 | 0.139 |
| 107 | 274 | 167 | 8.97 | 0.781 | 1.477 | 2.774 | 2.768 | 1.425 | 1.419 | 0.787 | 0.781 | 1.414 | 1.649 | 0.394 | 0.572 |
| 108 | 264 | 156 | 10.59 | 0.864 | -2.796 | -0.603 | -0.611 | -1.903 | -1.911 | -2.478 | -2.484 | -1.908 | -1.668 | -2.899 | -2.713 |
| 108 | 265 | 157 | 10.47 | 0.857 | -2.708 | -0.423 | -0.431 | -1.729 | -1.736 | -2.308 | -2.315 | -1.734 | -1.494 | -2.727 | -2.541 |
| 108 | 267 | 159 | 10.037 | 0.837 | -1.187 | 0.503 | 0.495 | -0.820 | -0.827 | -1.424 | -1.431 | -0.829 | -0.588 | -1.835 | -1.650 |
| 108 | 268 | 160 | 9.62 | 0.818 | 0.152 | 1.592 | 1.585 | 0.252 | 0.245 | -0.379 | -0.385 | 0.238 | 0.482 | -0.782 | -0.597 |
| 108 | 269 | 161 | 9.32 | 0.804 | 1.182 | 2.404 | 2.397 | 1.051 | 1.045 | 0.401 | 0.396 | 1.035 | 1.279 | 0.005 | 0.190 |
| 108 | 270 | 162 | 9.15 | 0.795 | 0.881 | 2.848 | 2.841 | 1.489 | 1.482 | 0.830 | 0.824 | 1.472 | 1.716 | 0.437 | 0.621 |
| 108 | 273 | 165 | 9.73 | 0.815 | -0.041 | 0.842 | 0.835 | -0.491 | -0.497 | -1.098 | -1.104 | -0.497 | -0.261 | -1.502 | -1.321 |
| 108 | 275 | 167 | 9.44 | 0.800 | -0.538 | 1.589 | 1.582 | 0.245 | 0.238 | -0.377 | -0.383 | 0.237 | 0.472 | -0.775 | -0.595 |
| 109 | 268 | 159 | 10.67 | 0.860 | -1.678 | -0.861 | -0.869 | -2.171 | -2.179 | -2.748 | -2.755 | -2.176 | -1.935 | -3.166 | -2.980 |
| 109 | 274 | 165 | 10.04 | 0.826 | -0.353 | 0.250 | 0.243 | -1.084 | -1.091 | -1.687 | -1.693 | -1.091 | -0.852 | -2.093 | -1.910 |
| 109 | 275 | 166 | 10.48 | 0.842 | -2.013 | -1.011 | -1.019 | -2.326 | -2.333 | -2.895 | -2.901 | -2.325 | -2.092 | -3.308 | -3.127 |
| 109 | 276 | 167 | 9.81 | 0.813 | -0.143 | 0.775 | 0.768 | -0.568 | -0.575 | -1.182 | -1.188 | -0.575 | 0.338 | -1.582 | -1.401 |
| 109 | 278 | 169 | 9.59 | 0.802 | 0.556 | 1.324 | 1.317 | -0.028 | -0.035 | -0.651 | -0.657 | -0.036 | 0.201 | -1.047 | -0.866 |
| 110 | 267 | 157 | 11.78 | 0.905 | -5 | -2.928 | -2.936 | -4.206 | -4.214 | -4.730 | -4.737 | -4.201 | -3.963 | -5.163 | -4.977 |
| 110 | 269 | 159 | 11.51 | 0.891 | -3.747 | -2.585 | -2.594 | -3.874 | -3.882 | -4.410 | -4.417 | -3.871 | -3.633 | -4.839 | -4.653 |
| 110 | 270 | 160 | 11.12 | 0.875 | -3.688 | -1.790 | -1.798 | -3.095 | -3.103 | -3.655 | -3.662 | -3.096 | -2.856 | -4.077 | -3.891 |
| 110 | 271 | 161 | 10.87 | 0.863 | -2.788 | -1.289 | -1.297 | -2.605 | -2.612 | -3.179 | -3.186 | -2.608 | -2.367 | -3.596 | -3.410 |
| 110 | 273 | 163 | 11.37 | 0.879 | -3.77 | -2.699 | -2.708 | -3.992 | -4.000 | -4.525 | -4.532 | -3.986 | -3.752 | -4.951 | -4.768 |
| 110 | 277 | 167 | 10.83 | 0.852 | -2.387 | -1.711 | -1.719 | -3.025 | -3.032 | -3.584 | -3.590 | -3.022 | -2.788 | -3.999 | -3.817 |
| 110 | 279 | 169 | 9.84 | 0.810 | 0.301 | 0.855 | 0.848 | -0.501 | -0.508 | -1.124 | -1.130 | -0.509 | -0.269 | -1.520 | 1.338 |
| 110 | 281 | 171 | 8.86 | 0.767 | 2.301 | 3.873 | 3.866 | 2.476 | 2.469 | 1.786 | 1.781 | 2.458 | 2.702 | 1.411 | 1.593 |
| 111 | 272 | 161 | 11.197 | 0.874 | -2.42 | -1.801 | -1.809 | -3.117 | -3.125 | -3.686 | -3.693 | -3.120 | -2.877 | -4.105 | -3.917 |
| 111 | 278 | 167 | 10.85 | 0.851 | -2.377 | -1.506 | -1.514 | -2.834 | -2.841 | -3.409 | -3.415 | -2.835 | -2.597 | -3.821 | -3.636 |

| 112 | 281 | 169 | 10.46 | 0.832 | -1 | -0.365 | -0.373 | -1.724 | -1.731 | -2.338 | -2.344 | -1.731 | -1.488 | -2.738 | -2.552 |
| 112 | 283 | 171 | 9.725 | 0.800 | 0.58 | 1.629 | 1.622 | 0.239 | 0.232 | -0.422 | -0.428 | 0.225 | 0.472 | -0.807 | -0.621 |
| 112 | 285 | 173 | 9.328 | 0.781 | 1.462 | 2.790 | 2.783 | 1.384 | 1.378 | 0.699 | 0.694 | 1.367 | 1.615 | 0.323 | 0.509 |
| 113 | 283 | 170 | 10.313 | 0.823 | -1 | 0.229 | 0.221 | -1.150 | -1.158 | -1.790 | -1.796 | -1.162 | -0.914 | -2.182 | -1.994 |
| 113 | 284 | 171 | 10.191 | 0.817 | -0.319 | 0.514 | 0.507 | -0.870 | -0.877 | -1.515 | -1.522 | -0.882 | -0.634 | -1.905 | -1.717 |
| 113 | 285 | 172 | 9.927 | 0.805 | 0.738 | 1.230 | 1.223 | -0.165 | -0.172 | -0.827 | -0.833 | -0.180 | 0.070 | -1.211 | -1.024 |
| 114 | 286 | 172 | 10.373 | 0.821 | -0.886 | 0.189 | 0.181 | -1.202 | -1.209 | -1.849 | -1.855 | -1.214 | -0.964 | -2.237 | -2.048 |
| 114 | 287 | 173 | 10.211 | 0.813 | -0.319 | 0.597 | 0.590 | -0.800 | -0.807 | -1.456 | -1.462 | -0.814 | -0.563 | -1.841 | -1.653 |
| 114 | 288 | 174 | 10.08 | 0.807 | -0.097 | 0.933 | 0.926 | -0.469 | -0.476 | -1.132 | -1.138 | -0.483 | -0.233 | -1.514 | -1.326 |
| 114 | 289 | 175 | 10.007 | 0.802 | 0.415 | 1.112 | 1.105 | -0.293 | -0.300 | -0.958 | -0.964 | -0.307 | -0.057 | -1.339 | -1.151 |
| 115 | 287 | 172 | 10.789 | 0.836 | -1.456 | -0.712 | -0.720 | -2.099 | -2.107 | -2.734 | -2.740 | -2.110 | -1.859 | -3.126 | -2.936 |
| 115 | 288 | 173 | 10.677 | 0.830 | -1.06 | -0.469 | -0.477 | -1.861 | -1.868 | -2.501 | -2.507 | -1.872 | -1.621 | -2.891 | -2.701 |
| 115 | 289 | 174 | 10.504 | 0.822 | -0.658 | -0.046 | -0.054 | -1.445 | -1.452 | -2.094 | -2.100 | -1.457 | -1.206 | -2.481 | -2.291 |
| 116 | 290 | 174 | 11.042 | 0.841 | -2.167 | -1.270 | -1.278 | -2.660 | -2.668 | -3.288 | -3.294 | -2.669 | -2.418 | -3.681 | -3.490 |
| 116 | 291 | 175 | 10.94 | 0.835 | -1.699 | -1.053 | -1.061 | -2.447 | -2.455 | -3.080 | -3.086 | -2.457 | -2.206 | -3.470 | -3.280 |
| 116 | 292 | 176 | 10.858 | 0.831 | -1.397 | -0.879 | -0.886 | -2.276 | -2.283 | -2.911 | -2.917 | -2.285 | -2.034 | -3.300 | -3.110 |
| 116 | 293 | 177 | 10.736 | 0.825 | -1.097 | -0.586 | -0.594 | -1.988 | -1.995 | -2.629 | -2.635 | -1.998 | -1.747 | -3.015 | -2.826 |
| 117 | 293 | 176 | 11.233 | 0.843 | -1.824 | -1.656 | -1.664 | -3.052 | -3.059 | -3.677 | -3.684 | -3.060 | -2.808 | -4.069 | -3.879 |

| Parent Nuclei | | | $Q_\alpha(Mev)$ | Temp (Mev) | $log_{10}T_{1/2}/s$ | | | | | |
|---|---|---|---|---|---|---|---|---|---|---|
| | | | | | Expt | Ngô80 | | Guo2013 | | Ni |
| Z | A | N | | | | T-IND | T-DEP | T-IND | T-DEP | |
| 104 | 259 | 155 | 9.13 | 0.811 | 0.415 | 1.679 | 1.691 | -0.098 | 0.044 | 0.147 |
| 104 | 263 | 159 | 8.25 | 0.766 | 3.301 | 4.249 | 4.261 | 2.311 | 2.451 | 2.932 |
| 105 | 256 | 151 | 9.34 | 0.825 | 0.453 | 2.000 | 2.013 | 0.112 | 0.258 | 0.284 |
| 105 | 258 | 153 | 9.5 | 0.829 | 0.747 | 1.267 | 1.280 | -0.649 | -0.503 | -0.167 |
| 105 | 259 | 154 | 9.62 | 0.832 | -0.292 | 0.801 | 0.815 | -1.125 | -0.978 | -1.049 |
| 105 | 270 | 165 | 8.02 | 0.745 | 3.556 | 5.251 | 5.265 | 2.937 | 3.082 | 4.524 |
| 105 | 263 | 158 | 8.83 | 0.792 | 1.798 | 2.812 | 2.825 | 0.728 | 0.874 | 1.258 |
| 106 | 259 | 153 | 9.8 | 0.840 | -0.507 | 0.879 | 0.894 | -1.228 | -1.077 | -1.065 |
| 106 | 261 | 155 | 9.71 | 0.833 | -0.728 | 0.890 | 0.906 | -1.269 | -1.118 | -0.820 |
| 106 | 263 | 157 | 9.4 | 0.816 | 0.033 | 1.582 | 1.598 | -0.650 | -0.499 | 0.047 |
| 106 | 267 | 161 | 8.32 | 0.763 | 2.803 | 4.806 | 4.821 | 2.377 | 2.526 | 3.435 |
| 106 | 269 | 163 | 8.7 | 0.777 | 2.681 | 3.353 | 3.369 | 0.918 | 1.068 | 2.173 |
| 106 | 271 | 165 | 8.66 | 0.772 | 2.212 | 3.397 | 3.413 | 0.912 | 1.062 | 2.303 |
| 107 | 260 | 153 | 10.4 | 0.863 | -1.456 | -0.275 | -0.257 | -2.548 | -2.392 | -1.888 |
| 107 | 261 | 154 | 10.5 | 0.865 | -1.928 | -0.663 | -0.645 | -2.949 | -2.793 | -2.686 |
| 107 | 264 | 157 | 9.96 | 0.838 | -0.357 | 0.421 | 0.440 | -1.985 | -1.829 | -0.758 |
| 107 | 265 | 158 | 9.38 | 0.812 | -0.027 | 2.003 | 2.021 | -0.485 | -0.330 | 0.298 |
| 107 | 266 | 159 | 9.43 | 0.813 | 0.398 | 1.758 | 1.776 | -0.748 | -0.593 | 0.706 |
| 107 | 267 | 160 | 8.96 | 0.791 | 1.342 | 3.148 | 3.166 | 0.570 | 0.724 | 1.559 |
| 107 | 270 | 163 | 9.06 | 0.791 | 1.778 | 2.611 | 2.629 | -0.026 | 0.129 | 1.804 |
| 107 | 272 | 165 | 9.14 | 0.792 | 0.913 | 2.252 | 2.270 | -0.423 | -0.268 | 1.563 |
| 107 | 274 | 167 | 8.97 | 0.781 | 1.477 | 2.725 | 2.744 | -0.015 | 0.140 | 2.083 |
| 108 | 264 | 156 | 10.59 | 0.864 | -2.796 | -0.695 | -0.674 | -3.249 | -3.088 | -3.197 |
| 108 | 265 | 157 | 10.47 | 0.857 | -2.708 | -0.502 | -0.481 | -3.092 | -2.931 | -2.155 |
| 108 | 267 | 159 | 10.037 | 0.837 | -1.187 | 0.451 | 0.471 | -2.232 | -2.072 | -1.045 |
| 108 | 268 | 160 | 9.62 | 0.818 | 0.152 | 1.552 | 1.573 | -1.199 | -1.039 | -0.653 |
| 108 | 269 | 161 | 9.32 | 0.804 | 1.182 | 2.378 | 2.399 | -0.430 | -0.271 | 0.960 |
| 108 | 270 | 162 | 9.15 | 0.795 | 0.881 | 2.838 | 2.858 | -0.013 | 0.146 | 0.723 |
| 108 | 273 | 165 | 9.73 | 0.815 | -0.041 | 0.882 | 0.904 | -1.976 | -1.816 | -0.211 |
| 108 | 275 | 167 | 9.44 | 0.800 | -0.538 | 1.660 | 1.682 | -1.279 | -1.119 | 0.612 |
| 109 | 268 | 159 | 10.67 | 0.860 | -1.678 | -0.794 | -0.771 | -3.654 | -3.489 | -1.928 |
| 109 | 274 | 165 | 10.04 | 0.826 | -0.353 | 0.413 | 0.437 | -2.666 | -2.501 | -0.320 |
| 109 | 275 | 166 | 10.48 | 0.842 | -2.013 | -0.833 | -0.808 | -3.885 | -3.719 | -2.008 |
| 109 | 276 | 167 | 9.81 | 0.813 | -0.143 | 0.972 | 0.996 | -2.184 | -2.019 | 0.305 |
| 109 | 278 | 169 | 9.59 | 0.802 | 0.556 | 1.556 | 1.580 | -1.677 | -1.512 | 0.924 |
| 110 | 267 | 157 | 11.78 | 0.905 | -5 | -2.773 | -2.746 | -5.710 | -5.540 | -4.560 |
| 110 | 269 | 159 | 11.51 | 0.891 | -3.747 | -2.399 | -2.372 | -5.416 | -5.246 | -3.977 |
| 110 | 270 | 160 | 11.12 | 0.875 | -3.688 | -1.587 | -1.560 | -4.673 | -4.504 | -3.845 |
| 110 | 271 | 161 | 10.87 | 0.863 | -2.788 | -1.068 | -1.042 | -4.210 | -4.040 | -2.511 |
| 110 | 273 | 163 | 11.37 | 0.879 | -3.77 | -2.448 | -2.420 | -5.580 | -5.410 | -3.665 |
| 110 | 277 | 167 | 10.83 | 0.852 | -2.387 | -1.389 | -1.361 | -4.687 | -4.517 | -2.412 |
| 110 | 279 | 169 | 9.84 | 0.810 | 0.301 | 1.223 | 1.251 | -2.251 | -2.082 | 0.145 |
| 110 | 281 | 171 | 8.86 | 0.767 | 2.301 | 4.296 | 4.323 | 0.636 | 0.802 | 3.089 |
| 111 | 272 | 161 | 11.197 | 0.874 | -2.42 | -1.450 | -1.420 | -4.818 | -4.644 | -2.570 |
| 111 | 278 | 167 | 10.85 | 0.851 | -2.377 | -1.045 | -1.013 | -4.615 | -4.440 | -1.748 |
| 112 | 281 | 169 | 10.46 | 0.832 | -1 | 0.301 | 0.338 | -3.672 | -3.494 | -0.869 |
| 112 | 283 | 171 | 9.725 | 0.800 | 0.58 | 2.366 | 2.403 | -1.781 | -1.605 | 1.135 |
| 112 | 285 | 173 | 9.328 | 0.781 | 1.462 | 3.592 | 3.630 | -0.684 | -0.508 | 2.315 |
| 113 | 283 | 170 | 10.313 | 0.823 | -1 | 1.108 | 1.151 | -3.242 | -3.060 | -0.310 |
| 113 | 284 | 171 | 10.191 | 0.817 | -0.319 | 1.424 | 1.731 | -2.980 | -2.798 | 0.569 |
| 113 | 285 | 172 | 9.927 | 0.805 | 0.738 | 2.410 | 2.410 | -2.304 | -2.123 | 0.747 |
| 114 | 286 | 172 | 10.373 | 0.821 | -0.886 | 1.417 | 1.429 | -3.432 | -3.246 | -0.751 |
| 114 | 287 | 173 | 10.211 | 0.813 | -0.319 | 1.821 | 1.878 | -3.052 | -2.866 | 0.432 |
| 114 | 288 | 174 | 10.08 | 0.807 | -0.097 | 2.200 | 0.260 | -2.740 | -2.555 | 0.046 |
| 114 | 289 | 175 | 10.007 | 0.802 | 0.415 | 2.419 | 2.474 | -2.579 | -2.393 | 0.997 |
| 115 | 287 | 172 | 10.789 | 0.836 | -1.456 | 0.696 | 0.755 | -4.425 | -4.234 | -0.914 |
| 115 | 288 | 173 | 10.677 | 0.830 | -1.06 | 0.978 | 1.034 | -4.204 | -4.014 | -0.076 |
| 115 | 289 | 174 | 10.504 | 0.822 | -0.658 | 1.443 | 1.496 | -3.811 | -3.620 | -0.177 |
| 116 | 290 | 174 | 11.042 | 0.841 | -2.167 | 0.411 | 0.465 | -5.113 | -4.918 | -1.839 |
| 116 | 291 | 175 | 10.94 | 0.835 | -1.699 | 0.659 | 0.711 | -4.917 | -4.722 | -0.838 |

| 116 | 292 | 176 | 10.858 | 0.831 | -1.397 | 0.864 | 0.915 | -4.761 | -4.566 | -1.379 |
| 116 | 293 | 177 | 10.736 | 0.825 | -1.097 | 1.188 | 1.237 | -4.492 | -4.297 | -0.320 |
| 117 | 293 | 176 | 11.233 | 0.843 | -1.824 | 0.277 | 0.328 | -5.638 | -5.438 | -1.399 |

To intuitively observe the deviations between the calculated α-decay half-lives of superheavy nuclei and the experimental half-lives, this study employs the root-mean-square deviation ($\sigma$) as a metric. It can be expressed as follows:

$$\sigma = \sqrt{\sum_{i=1}^{n} \frac{\left(log_{10} T_{1/2}^{exp,i} - log_{10} T_{1/2}^{cal,i}\right)^2}{n}}. \#(34)$$

In this context, $log_{10} T_{1/2}^{exp,i}$ and $log_{10} T_{1/2}^{cal,i}$ represent the logarithmic forms of the experimental and calculated α-decay half-lives for the i-th nucleus, respectively. $n$ denotes the number of nuclei involved in different decay scenarios. Table 3 presents the detailed calculations of the root-mean-square deviations ($\sigma$) for 22 different versions of proximity potentials, including both temperature-independent and temperature-dependent forms. Additionally, the root-mean-square deviations ($\sigma$) obtained by using Ni's empirical formula are also listed in the table. From this table, it is evident that for proximity potentials such as Prox.77-3, Prox.77-6, Prox.77-7, Bass73, Bass77, AW95, and Guo2013, the root-mean-square deviations ($\sigma$) decrease when temperature effects are incorporated. This indicates that incorporating temperature dependence in these proximity potentials improves the accuracy of calculating the α-decay half-lives of superheavy nuclei. Furthermore, it can be observed that the most suitable proximity potential for describing α-decay in superheavy nuclei is Prox.77-3 T-DEP, as it exhibits the lowest root-mean-square deviation of $\sigma$ = 0.515. Other proximity potentials with root-mean-square deviations ($\sigma$) less than 1 can also be applied to describe α-decay in superheavy nuclei. However, some proximity potentials, such as Bass77, AW95, Ngô80, and Guo2013, exhibit significant deviations between their calculated results and the experimental data. Overall, the results obtained from the CPPM model by using various proximity potential formalisms demonstrate comparable levels of accuracy when compared to experimental data. To further validate the feasibility of applying proximity potential to α-decay in superheavy nuclei, we plot the differences between the experimentally measured α-decay half-lives of superheavy nuclei and those calculated using the CPPM model combined with Prox.77-3 T-DEP, Ngô T-IND, and Guo2013 T-IND in logarithmic form in Fig. 1. These results are also compared with those obtained by using Ni's empirical formulas. From the figure, it is evident that the deviations calculated using the CPPM model combined with Prox.77-3 T-DEP proximity potential and Ni's empirical formula are primarily distributed within the range of -1 to 1, while the results obtained by using the CPPM model combined with the Ngô T-IND and Guo2013 T-IND proximity potential exhibit relatively larger dispersion. This demonstrates that, among the 22 proximity potential formalisms, Prox.77-3 T-DEP provide the most accurate theoretical predictions for the α-decay half-lives of superheavy nuclei.

Table3. The root-mean-square deviation ($\sigma$) between the experimental data and the calculated results for α-decay of superheavy nucleiwas was obtained using the CPPM model combined with 22 proximity potential formalisms, each including both temperature-dependent and temperature-independent cases, as well as Ni's empirical formula.

| method | Prox.77-1 | | Prox.77-2 | | Prox.77-3 | | Prox.77-4 | | Prox.77-5 | | Prox.77-6 | | Prox.77-7 | |
|---|---|---|---|---|---|---|---|---|---|---|---|---|---|---|
| | T-IND | T-DEP | T-IND | T-DEP | T-IND | T-DEP | T-IND | T-DEP | T-IND | T-DEP | T-IND | T-DEP | T-IND | T-DEP |
| $\sigma$ | 0.642 | 0.851 | 0.568 | 0.718 | 0.580 | 0.515 | 0.522 | 0.540 | 0.521 | 0.555 | 0.541 | 0.519 | 0.535 | 0.524 |
| method | Prox.77-8 | | Prox.77-9 | | Prox.77-10 | | Prox.77-11 | | Prox.77-12 | | Prox.77-13 | | Prox.81 | |
| | T-IND | T-DEP | T-IND | T-DEP | T-IND | T-DEP | T-IND | T-DEP | T-IND | T-DEP | T-IND | T-DEP | T-IND | T-DEP |
| $\sigma$ | 0.695 | 0.936 | 0.537 | 0.701 | 0.537 | 0.700 | 0.537 | 0.625 | 0.634 | 0.836 | 0.940 | 1.016 | 0.649 | 0.870 |
| method | Bass73 | | Bass77 | | Bass80 | | CW76 | | BW91 | | AW95 | | Ngô80 | |
| | T-IND | T-DEP | T-IND | T-DEP | T-IND | T-DEP | T-IND | T-DEP | T-IND | T-DEP | T-IND | T-DEP | T-IND | T-DEP |
| $\sigma$ | 0.622 | 0.619 | 1.345 | 1.338 | 0.564 | 0.566 | 0.910 | 0.915 | 0.565 | 0.569 | 1.246 | 1.072 | 1.645 | 1.653 |
| method | Guo2013 | | Ni | | | | | | | | | | | |
| | T-IND | T-DEP | | | | | | | | | | | | |
| $\sigma$ | 1.966 | 1.814 | 0.509 | | | | | | | | | | | |

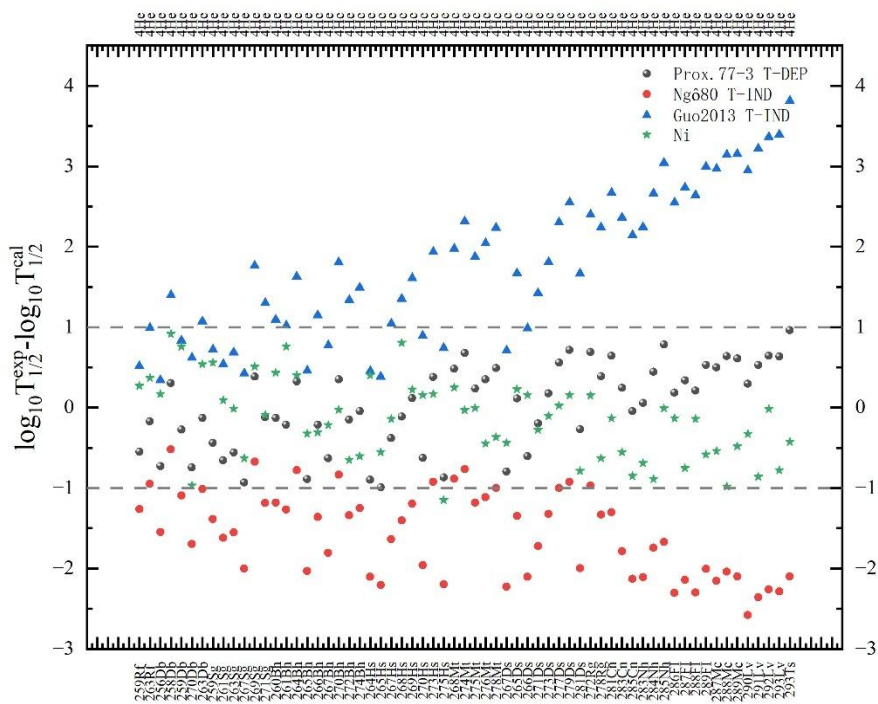

Figure 1 (Color online) Comparison of the discrepancy between the experimentally measured $\alpha$-decay half-lives of superheavy nuclei and the calculated results obtained using the CPPM

model combined with the Prox.77-3 T-DEP, Ngô80 T-IND, and Guo2013 T-IND proximity potentials, as well as the Ni's empirical formula, presented in logarithmic form.

Motivated by the excellent agreement between the experimentally measured α-decay half-lives of superheavy nuclei and the calculated results obtained by using the CPPM model combined with the Prox.77-3 T-DEP proximity potential, we further selected five proximity potentials with the smallest root-mean-square deviations to theoretically predict the α-decay half-lives of 36 potential superheavy nuclei. The predicted results are listed in Table 4. In this table, the first three columns show the proton number (Z), mass number (A), and neutron number (N) of the parent nuclei, respectively. The fourth and fifth columns provide the α-decay energy (Q) and the temperature (T) of the parent nuclei, respectively. The last six columns present the predicted results in logarithmic form, calculated using the CPPM model combined with the proximity potentials Prox.77-13 T-DEP, Prox.77-4 T-IND, Prox.77-5 T-IND, Prox.77-6 T-DEP, and Prox.77-7 T-DEP, as well as the Ni's empirical formula. From Table 4, it is evident that the predicted results obtained by using the CPPM model combined with the five proximity potentials and the Ni's empirical formula exhibit consistency in the same order of magnitude for the α-decay half-lives of superheavy nuclei and agree well with the experimental data. To visually compare the consistency between our predictions and the Ni's empirical formula for α-decay in superheavy nuclei, we plot the logarithmic half-lives calculated using the CPPM model combined with the proximity potential yielding the smallest root-mean-square deviation and the Ni's empirical formula in Figure 3. From the figure, it is evident that the results obtained using the CPPM model combined with the five proximity potentials are in excellent agreement with those from the Ni's empirical formula, demonstrating the high reliability of the CPPM model incorporating proximity potentials in the theoretical calculation of α-decay half-lives for superheavy nuclei. We anticipate that these predictions will provide valuable theoretical references for forthcoming experimental investigations into the alpha decay of superheavy nuclei.

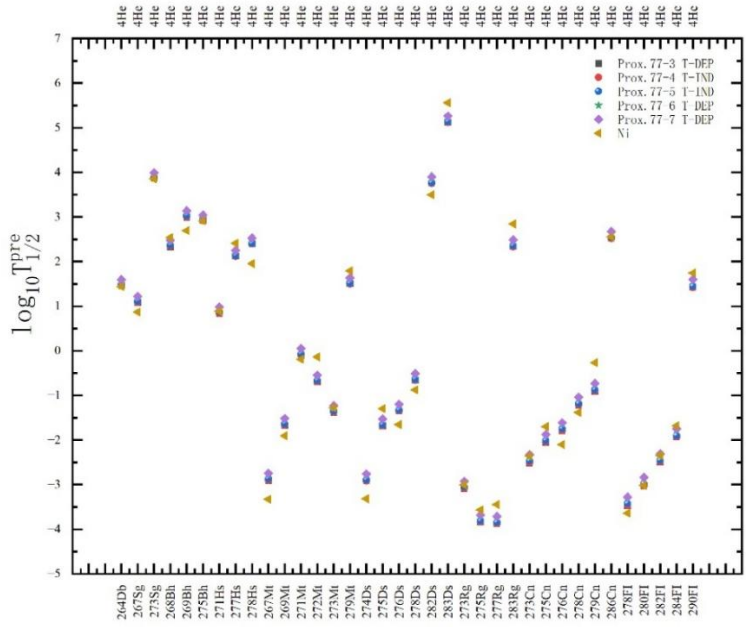

Figure 2 (Color online) Predicted α-decay half-lives of superheavy nuclei obtained using the CPPM model combined with the Prox.77-3 T-DEP, Prox.77-4 T-IND, Prox.77-5 T-IND, Prox.77-6 T-DEP, and Prox.77-7 T-DEP proximity potentials, as well as the Ni empirical formula, presented in logarithmic form.

Table4. The Predicted half-lives for 36 potential superheavy nuclei undergoing $\alpha$-decay are presented, with the $Q_\alpha$ values taken from Refs [27,30].

| Parent Nuclei | | | $Q_\alpha$(Mev) | Temp(Mev) | $log_{10}T_{1/2}/s$ | | | | | |
|---|---|---|---|---|---|---|---|---|---|---|
| | | | | | Prox.77-3 | Prox.77-4 | Prox.77-5 | Prox.77-6 | Prox.77-7 | Niox.77-1 |
| Z | A | N | | | T-DEP | T-IND | T-IND | T-DEP | T-DEP | |
| 105 | 264 | 159 | 8.95 | 0.795 | 1.461 | 1.470 | 1.497 | 1.573 | 1.598 | 1.441 |
| 106 | 267 | 161 | 9.12 | 0.798 | 1.081 | 1.088 | 1.115 | 1.194 | 1.219 | 0.868 |
| 106 | 273 | 167 | 8.2 | 0.749 | 3.872 | 3.871 | 3.885 | 3.966 | 3.991 | 3.855 |
| 107 | 268 | 161 | 8.824 | 0.784 | 2.328 | 2.342 | 2.374 | 2.453 | 2.480 | 2.539 |
| 107 | 269 | 162 | 8.605 | 0.773 | 2.987 | 3.001 | 3.030 | 3.110 | 3.137 | 2.697 |
| 107 | 275 | 168 | 8.541 | 0.762 | 2.919 | 2.916 | 2.932 | 3.017 | 3.042 | 2.912 |
| 108 | 271 | 163 | 9.346 | 0.802 | 0.830 | 0.838 | 0.868 | 0.954 | 0.980 | 0.885 |
| 108 | 277 | 169 | 8.85 | 0.772 | 2.120 | 2.115 | 2.133 | 2.222 | 2.248 | 2.408 |
| 108 | 278 | 170 | 8.76 | 0.767 | 2.401 | 2.394 | 2.409 | 2.499 | 2.525 | 1.952 |
| 109 | 267 | 158 | 11.027 | 0.876 | -2.911 | -2.903 | -2.863 | -2.774 | -2.748 | -3.328 |
| 109 | 269 | 160 | 10.438 | 0.850 | -1.680 | -1.672 | -1.634 | -1.544 | -1.518 | -1.905 |
| 109 | 271 | 162 | 9.789 | 0.820 | -0.110 | -0.100 | -0.065 | 0.024 | 0.052 | -0.190 |
| 109 | 272 | 163 | 9.972 | 0.826 | -0.704 | -0.699 | -0.665 | -0.575 | -0.548 | -0.139 |
| 109 | 273 | 164 | 10.194 | 0.833 | -1.380 | -1.379 | -1.349 | -1.256 | -1.230 | -1.278 |
| 109 | 279 | 170 | 9.113 | 0.780 | 1.502 | 1.496 | 1.515 | 1.609 | 1.636 | 1.790 |
| 110 | 274 | 164 | 10.896 | 0.860 | -2.915 | -2.917 | -2.883 | -2.787 | -2.760 | -3.318 |

| | | | | | | | | | |
|---|---|---|---|---|---|---|---|---|---|
| 110 | 275 | 165 | 10.38 | 0.838 | -1.686 | -1.686 | -1.653 | -1.557 | -1.530 | -1.296 |
| 110 | 276 | 166 | 10.23 | 0.830 | -1.352 | -1.353 | -1.322 | -1.225 | -1.198 | -1.654 |
| 110 | 278 | 168 | 9.94 | 0.816 | -0.659 | -0.663 | -0.636 | -0.539 | -0.511 | -0.877 |
| 110 | 282 | 172 | 8.515 | 0.751 | 3.755 | 3.753 | 3.773 | 3.868 | 3.897 | 3.498 |
| 110 | 283 | 173 | 8.146 | 0.733 | 5.119 | 5.118 | 5.136 | 5.230 | 5.260 | 5.561 |
| 111 | 273 | 162 | 11.148 | 0.871 | -3.095 | -3.091 | -3.051 | -2.952 | -2.924 | -3.008 |
| 111 | 275 | 164 | 11.393 | 0.877 | -3.843 | -3.846 | -3.810 | -3.709 | -3.682 | -3.567 |
| 111 | 277 | 166 | 11.34 | 0.872 | -3.871 | -3.879 | -3.846 | -3.743 | -3.717 | -3.447 |
| 111 | 283 | 172 | 9.002 | 0.770 | 2.334 | 2.332 | 2.356 | 2.456 | 2.486 | 2.842 |
| 112 | 273 | 161 | 11.06 | 0.867 | -2.519 | -2.509 | -2.462 | -2.361 | -2.332 | -2.356 |
| 112 | 275 | 163 | 10.79 | 0.854 | -2.061 | -2.053 | -2.009 | -1.906 | -1.877 | -1.703 |
| 112 | 276 | 164 | 10.65 | 0.847 | -1.797 | -1.790 | -1.747 | -1.644 | -1.615 | -2.101 |
| 112 | 278 | 166 | 10.37 | 0.833 | -1.222 | -1.217 | -1.178 | -1.074 | -1.044 | -1.383 |
| 112 | 279 | 167 | 10.23 | 0.826 | -0.910 | -0.907 | -0.869 | -0.765 | -0.735 | -0.266 |
| 112 | 286 | 174 | 9.014 | 0.766 | 2.520 | 2.515 | 2.539 | 2.645 | 2.676 | 2.555 |
| 114 | 278 | 164 | 11.55 | 0.878 | -3.478 | -3.474 | -3.425 | -3.312 | -3.281 | -3.642 |
| 114 | 280 | 166 | 11.28 | 0.865 | -3.029 | -3.027 | -2.981 | -2.867 | -2.836 | -3.019 |
| 114 | 282 | 168 | 11 | 0.851 | -2.499 | -2.500 | -2.457 | -2.341 | -2.309 | -2.349 |
| 114 | 284 | 170 | 10.73 | 0.838 | -1.934 | -1.938 | -1.898 | -1.781 | -1.749 | -1.678 |
| 114 | 290 | 176 | 9.495 | 0.781 | 1.426 | 1.417 | 1.446 | 1.565 | 1.598 | 1.746 |

## IV. SUMMARY

We have Systematically evaluated 22 proximity potentials, incorporating temperature dependence and diffusion parameters related to proton (Z) and neutron (N) numbers, to describe α-decay in superheavy nuclei using the Coulomb and Proximity Potential Model (CPPM). The results demonstrated that Prox.77-3 T-DEP achieved the highest accuracy, with the lowest root-mean-square deviation ($\sigma = 0.515$), closely matching experimental data. In contrast, proximity potentials such as Bass77, AW95, Ng ô 80, and Guo2013 exhibited significant deviations. Temperature dependence improved the precision of several potentials, including Prox.77-3, Prox.77-6, and Prox.77-7. Predictions for 36 potential superheavy nuclei were made, providing valuable theoretical insights for future experiments. Overall, this work highlights the effectiveness of CPPM combined with temperature-dependent proximity potentials in accurately modeling α-decay half-lives, offering a reliable framework for superheavy nuclear research.

## V. ACKNOWLEDGEMENTS

This work is supported by Yunnan Provincial Science Foundation Project (No. 202501AT070067), Yunnan Provincial Xing Dian Talent Support Program (Young Talents Special Program), Kunming University Talent Introduction Research Project (No. YJL24019), Yunnan Provincial Department of Education Scientific Research Fund Project (No. 2025Y1042 and 2025Y1055), the Program for Frontier Research